\documentclass[12pt,preprint]{aastex}
\usepackage{longtable}
\usepackage{pdflscape}	
\usepackage{multirow}
\usepackage{amsmath}
\usepackage{url}
\begin{document}
\title{The first photometric and spectroscopic analysis of the extremely low mass ratio contact binary NSVS 5029961}
\author{Shu-Yue Zheng\altaffilmark{1}, Kai Li\altaffilmark{1}, Qi-Qi Xia\altaffilmark{1}}

\altaffiltext{1}{Shandong Key Laboratory of Optical Astronomy and Solar-Terrestrial Environment, School of Space Science and Physics, Institute of Space Sciences, Shandong University, Weihai, Shandong, 264209, China (e-mail: kaili@sdu.edu.cn (KL))}

\begin{abstract}
We performed photometric and spectroscopic investigations of NSVS 5029961 for the first time. The new BV(RI)$_c$-band light curves were obtained with the 1.0-m telescope at Weihai Observatory of Shandong University. Applying the Wilson-Devinney program, we found that NSVS 5029961 is an A-subtype shallow contact binary with extremely low mass ratio (q = 0.1515, f = 19.1\%). Six spectra have been obtained by LAMOST, and many chromospheric activity emission line indicators were detected in the spectra, revealing that the target exhibits strong chromospheric activity. We calculated the absolute parameters with the photometric solutions and Gaia distance, and estimated the initial masses of the two components and the age of the binary. The evolutionary status was discussed by using the mass-radius and mass-luminosity diagrams. The result shows the primary component is a little evolved star and the secondary component has evolved away from the main sequence. The formation and evolution investigations of NSVS 5029661 indicate that it may have evolved from a detached binary with short period and low mass ratio by angular momentum loss via magnetic braking and case A mass transfer, and is in a stable contact stage at present.
\end{abstract}

\keywords{stars: binaries: close ---
         stars: binaries: eclipsing ---
         stars: evolution ---
         stars: individual (NSVS 5029961) }

\section{Introduction}
W UMa-type contact binary is generally composed of F, G, and K spectral type stars (Rucinski 1993) which fill the Roche-lobe (Kopal 1959) and have nearly same surface temperatures. The common convection envelope (CCE) shared by two components (Lucy 1968) is located between the inner and outer critical equipotential surfaces, where the mass and energy transfers occur frequently. The light curves of the systems with short orbital period (P $<$ 1 day) are usually classified as EW-type. W UMa binaries are divided into two subtypes: A and W (Binnendijk 1970). The hotter component of A-subtype binary has more mass, while the one of W-subtype binary has less mass. The formation and evolution of W UMa stars are unsolved at present. It is predicted that they result from cool detached binaries with short period via angular momentum loss (AML) controlled by magnetic activities (e.g., Huang 1966; van't Veer 1979; Vilhu 1982; Qian 2003; Li et al. 2007; Qian et al. 2017; Qian et al. 2018) and evolve into extreme low mass-ratio contact configurations (e.g., Li et al. 2004; Qian et al. 2005; Qian et al. 2006). Contact binaries are supposed to have a low mass ratio limit (e.g., Jiang et al. 2010; Yang \& Qian 2015; Wadhwa et al. 2021). When the mass ratio is smaller than the q$_{min}$, the system will merge into a fast-rotating single object (FK Com-type star or blue straggler) ultimately based on tidal instability (Webbink 1976; Hut 1980; Eggleton \& Kiseleva-Eggleton 2001). The extremely low mass-ratio systems are considered as the late stage of evolution and the progenitor candidates of luminous red novae (e.g., Zhu et al. 2016; Molnar et al. 2017), so their discovery and research are very valuable, which attracts the attention of more and more researchers (e.g., Kjurkchieva et al. 2018a; Sarotsakulchai et al. 2018; Zhou et al. 2020). V1309 Sco is an evidence of the red nova eruption from the coalescence of contact binary (Tylenda et al. 2011). Wadhwa et al. (2021) gave more reliable criteria of instability mass ratio and instability separation which depend on the mass of the primary star, corresponding gyration radii, the degree of contact and other physical parameters. New merger candidates were found with the method by them.

The strong chromosphere, transition region and coronal activities usually exist in late type systems (e.g., Eker et al. 2008). The magnetic activity may be connected with convective motions and rapid stellar rotation (e.g., Pavlenko et al. 2018; $\c{S}$enavci et al. 2018) and is presented as star spots, flares, and plages (e.g., Plibulla et al. 2003; Strassmeier 2009). Emission lines of Ca II H and K, Balmer series and Ca II triplet above the continuum, the indicators of chromospheric activity, were found in many contact binaries (e.g., Pavlenko et al. 2018; Long et al. 2019; Cheng et al. 2019; Zhang et al. 2020). The O'Connell effect (O'Connell 1951), the phenomenon that two maxima of the light curve are unequal, is a common characteristic of many contact binaries (e.g., Zhou et al. 2016; Zhou \& Soonthornthum 2019; Hu et al. 2020). Many researchers suspect that it is connected with magnetic-activity starspots (e.g., Binnendijk 1960; Qian et al. 2013; Lu et al. 2020). In addition to the formation, evolution and magnetic activity of W UMa binary systems, many other issues also need to be addressed such as the thermal relaxation oscillation (TRO) theory (e.g., Lucy 1976; Flannery 1976), AML (e.g., van't Veer 1979; Rahunen 1981) and period cut-off (e.g., Rucinski 1992; Weldrake et al. 2004; Li et al. 2019b), so more observations of contact binaries are needed. The precise mass ratio can be determined from the photometric light curves only for total eclipsing contact binaries (Pribulla et al. 2003; Terrell \& Wilson 2005). Therefore, the photometric solutions and high-precision distances given by Gaia mission (Gaia Collaboration et al. 2018; Bailer-Jones et al. 2018) provide the opportunity to obtain the reliable global parameters for total eclipsing contact binaries (e.g., Li et al. 2019a, 2021; Kjurkchieva et al. 2019a), which will help us study W UMa systems well.

NSVS 5029961 (ASASSN-V J123609.02+421435.1, TYC 3020-2405-1) was first identified as an EW type binaries by Sergey, Ivan  in 2014 (VSX database\nolinebreak\footnotemark[1]\footnotetext[1]{https://www.aavso.org/vsx/index.php?view=detail.top\&oid=399911}) and confirmed by All-Sky Automated Survey for Supernovae (ASAS-SN, Shappee et al. 2014; Jayasinghe et al. 2019). The orbit period is 0.3766626 days and the V-band mean magnitude is 11.62 mag with an amplitude of 0.29. After its discovery, no further study was published. In this paper, we presented multi-band photometric and spectroscopic analysis of NSVS 5029961.

\section{Observations and Orbital Period Analysis}
The photometric observations of NSVS 5029961 in BV(RI)$_c$ bands were carried out by using the 1.0-m telescope at Weihai Observatory of Shandong University (WHOT, Hu et al. 2014) on March 15, 2020. A back-illuminated PIXIS 2048B CCD detector with a field of view of $12^{'}\times12^{'}$ attached to the Cassegrain focus, and standard Johnson-Cousin-Bessel filters were adopted for the observations. Table 1 shows observing details on exposure times, observing errors, and the names, coordinates and magnitudes of the binary star, the comparison and the check stars. The photometric images were reduced by bias, dark and flat corrections and processed with aperture photometry method by using the C-Munipack\nolinebreak\footnotemark[2]\footnotetext[2]{http://c-munipack.sourceforge.net/} software. The differential magnitudes between the variable star and the comparison star and between the comparison and the check stars were determined, and the photometric data are listed in Table 2. Figure 1 displays the light curves which show the typical EW type. Flat eclipse can be seen in the secondary minimum, which lasts for about 40 minutes. Light curves of NSVS 5029961 were also found in TESS (Ricker et al. 2015), Wide Angle Search for Planets (SuperWASP, Butters et al. 2010), All-Sky Automated Survey for Supernovae (ASAS-SN, Shappee et al. 2014; Jayasinghe et al. 2019), Northern Sky Variability Survey (NSVS, Wozniak et al. 2004) and Catalina Sky Surveys (CSS, Drake et al. 2014). We divided the SuperWASP data into four parts by year. Only the two parts obtained in 2004 and 2007 were good enough for subsequent analysis. Then we deleted discrete points and binned the data to 200 points, respectively. The data obtained by CSS were too discrete for investigation. A total of five sets of light curves from the sky surveys were determined after the above processing and selection and are shown in Figure 2.

The spectroscopic observations of NSVS 5029961 were carried out by using LAMOST (The Large Sky Area Multi-Object Fiber Spectroscopic Telescope) from 2013 to 2017. LAMOST, a special reflecting Schmidt telescope with resolution of around 1800 for low resolution mode, has a field of view of $5^{\circ}$ and covers the wavelength from 3700{\AA} to 9000 {\AA} (Wang et al. 1996; Cui et al. 2012; Du et al. 2016). 4000 optical fibers were installed on the focal surface, improving the rate of spectral acquisition (Cui et al. 2012). We found six spectra with low resolution mode from Data Release 6\nolinebreak\footnotemark[3]\footnotetext[3]{http://dr6. lamost.org/} (He et al. 2017). The spectral parameters are tabulated in Table 3, including the observational date, Heliocentric Julian date, phase, exposure time, spectral type, the effective temperature, surface gravity and radial velocity.

As many times of minimum light as possible from our data and sky survey databases were searched for the investigation of orbital period change. Two eclipse times from our data and seventy-nine eclipse times from SuperWASP database were determined with the K-W method (Kwee \& van Woerden 1956). Although the observation cadence of TESS is 30 minutes, the data have high accuracy, so we tried to use them to derive more eclipse times by applying the method of Li et al. (2020). The light curve of TESS was divided into four parts. Each part was converted into one period by using the following equation,
\begin{equation}
BJD = BJD_0 + P\times E, \\
\end{equation}
where BJD is observing time, BJD$_0$ is the reference time and P is the orbital period. The corresponding four diagrams are plotted in Figure 3 and eight eclipse times were determined. We transformed the times of minimum light from HJD to BJD on the website\nolinebreak\footnotemark[4]\footnotetext[4]{http://astroutils.astronomy.ohio-state.edu/time/hjd2bjd.html} and lists all available minimum timings in Table 4. The data points of NSVS, CSS and ASAS-SN are sparse, so the times of minimum light from these databases cannot be obtained. O-C method was applied to study the orbital period variation. We also employed Equation (1) to calculate the values of O-C. Differently, BJD and BJD$_0$ represent the every eclipse time and the primary minima timing of our data respectively. The O-C diagram is presented in Figure 4. We adopted a linear equation to fit the points using least squares method. A new ephemeris was obtained:
\begin{equation}
Min.I = 2458924.04911(\pm0.00034) + 0.37666613(\pm0.00000003)\times E.\\
\end{equation}

\renewcommand\arraystretch{1.3}
\begin{table*}
\tiny
\begin{center}
\caption{Observing details of NSVS 5029961}
\begin{tabular}{p{1.94cm}p{2.2cm}p{1.3cm}p{1.3cm}p{0.8cm}p{0.8cm}p{0.8cm}p{1.75cm}p{1.84cm}}
\hline
Targets            & Name            & $\alpha_{2000}$  & $\delta _{2000}$    & mag    & mag     & mag      &Exposure times   &observing errors    \\
                   &                 &                  &                     &  (J)   &  (H)    & (K)      & (s)             &(mag)               \\
\hline
Binary star        & NSVS            & 12 36 08.94      & +42 14 35.0         & 10.643 & 10.396  & 10.352   & B50 V25         & B0.012 V0.009   \\
                   & 5029961         &                  &                     &        &         &          & $R_c$15 $I_c$13 & $R_c$0.007 $I_c$0.008\\
The comparison$^a$ & 2MASS           & 12 36 36.42 & +42 13 26.2 & 10.451 & 10.035 & 9.962 & & \\
               & 12363642+4213262& & & & &                                       & & \\
The check$^a$      & 2MASS & 12 36 37.61 & +42 11 38.6 & 11.829 & 11.466  & 11.338 & & \\
               & 12363761+4211386 & & & & & & &\\
\hline
\end{tabular}
\end{center}
$^a$ The names, coordinates and magnitudes of the comparison and check stars were determined from Two Micron All Sky Survey (2MASS; Cutri et al. 2003).
\end{table*}

\renewcommand\arraystretch{1.3}
\begin{table*}
\tiny
\begin{center}
\caption{The photometric data of NSVS 5029961 obtained on March 15, 2020}
\begin{tabular}{cccccccc}
\hline
HJD\_B &$\Delta$m\_B & HJD\_V &$\Delta$m\_V & HJD\_R &$\Delta$m\_R & HJD\_I &$\Delta$m\_I \\
(2400000+) &(mag)        &(2400000+)     &(mag)        &(2400000+)    &(mag)        & (2400000+)    &(mag) \\
\hline
58923.98445 &	-0.435 &	58923.98387 &	-0.254 &	58923.98352 &	-0.141 & 	58923.98322 &	-0.025\\
58923.98619 &	-0.411 &	58923.98561 &	-0.241 &	58923.98526 &	-0.142 &	58923.98496 &	-0.051\\
58923.98793 &	-0.412 &	58923.98735 &	-0.237 &	58923.98700 &	-0.133 &	58923.98670 &	-0.053\\
58923.98967 &	-0.412 &	58923.98909 &	-0.230 &	58923.98875 &	-0.139 &	58923.98844 &	-0.043\\
58923.99142 &	-0.396 &	58923.99084 &	-0.218 &	58923.99049 &	-0.131 &	58923.99018 &	-0.056\\
...          &...&...&...&...&...&...&...\\
58924.36460&-0.392&58924.36576&-0.199&58924.36541&-0.115&58924.36511&-0.025\\
58924.36634&-0.383&58924.36750&-0.198&58924.36715&-0.105&58924.36685&-0.022\\
58924.36808&-0.378&58924.36924&-0.201&58924.36890&-0.104&58924.36859&-0.021\\
\hline
\end{tabular}
\end{center}
This table is available in its entirety in machine-readable form. A portion is shown here for guidance regarding its
form and content.
\end{table*}

\begin{figure}
\begin{center}
\includegraphics[width=0.6\textwidth]{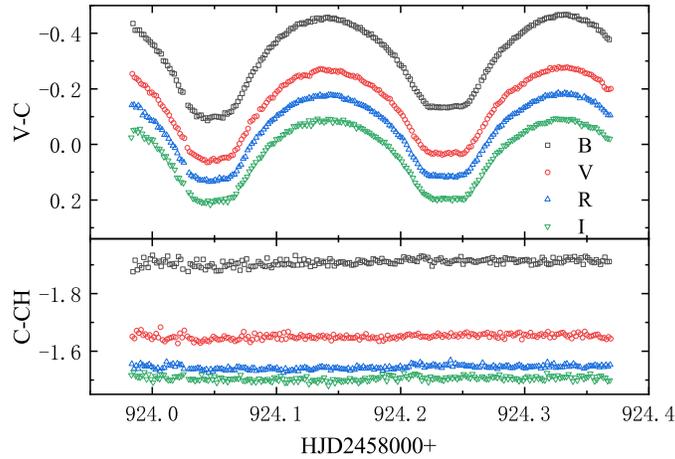}
\caption{Multi-color light curves of NSVS 5029961 obtained on March 15, 2020.}
\end{center}
\end{figure}

\begin{figure}
\begin{center}
\includegraphics[width=0.36\textwidth]{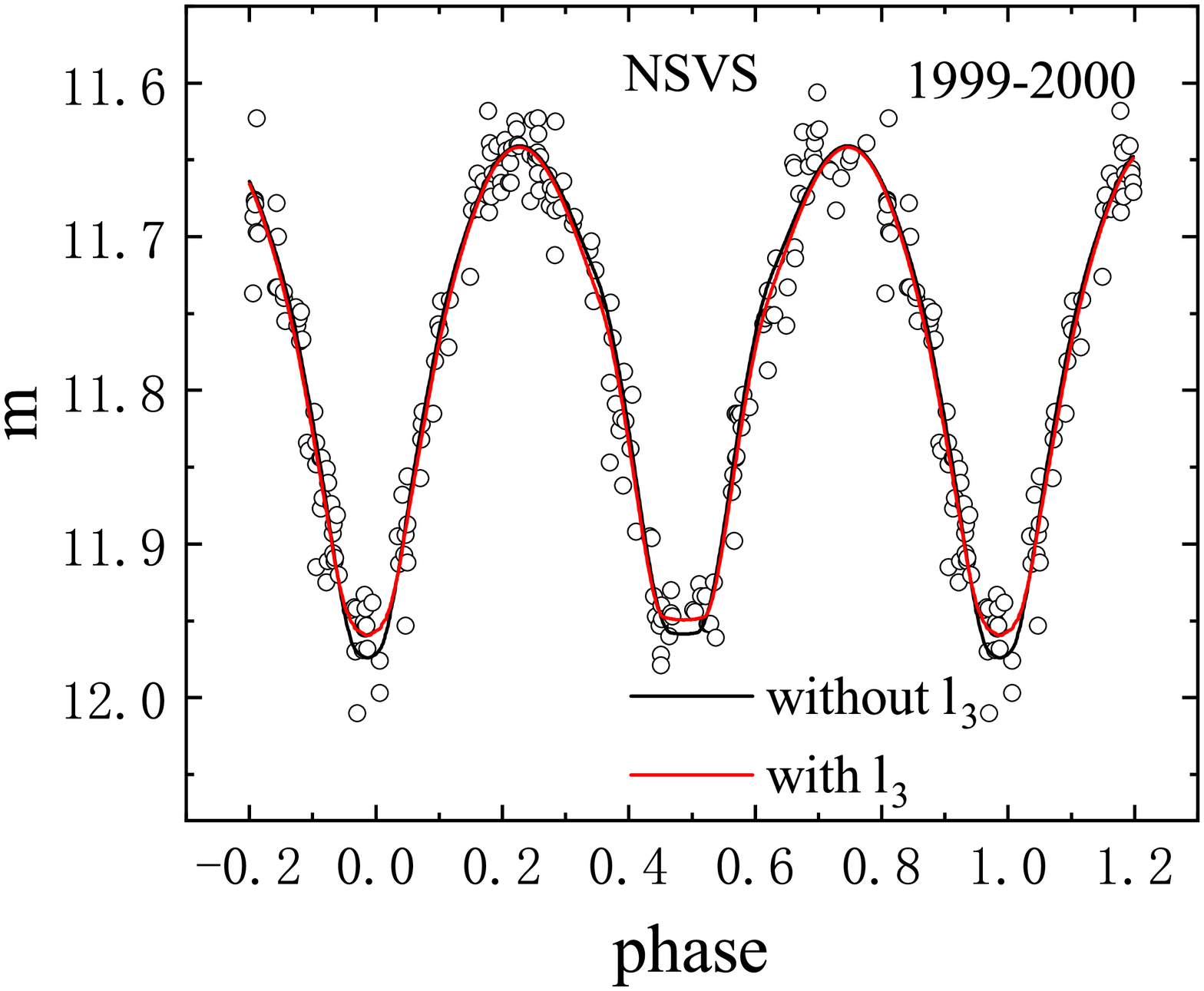}
\hspace{-1.1cm}
\includegraphics[width=0.36\textwidth]{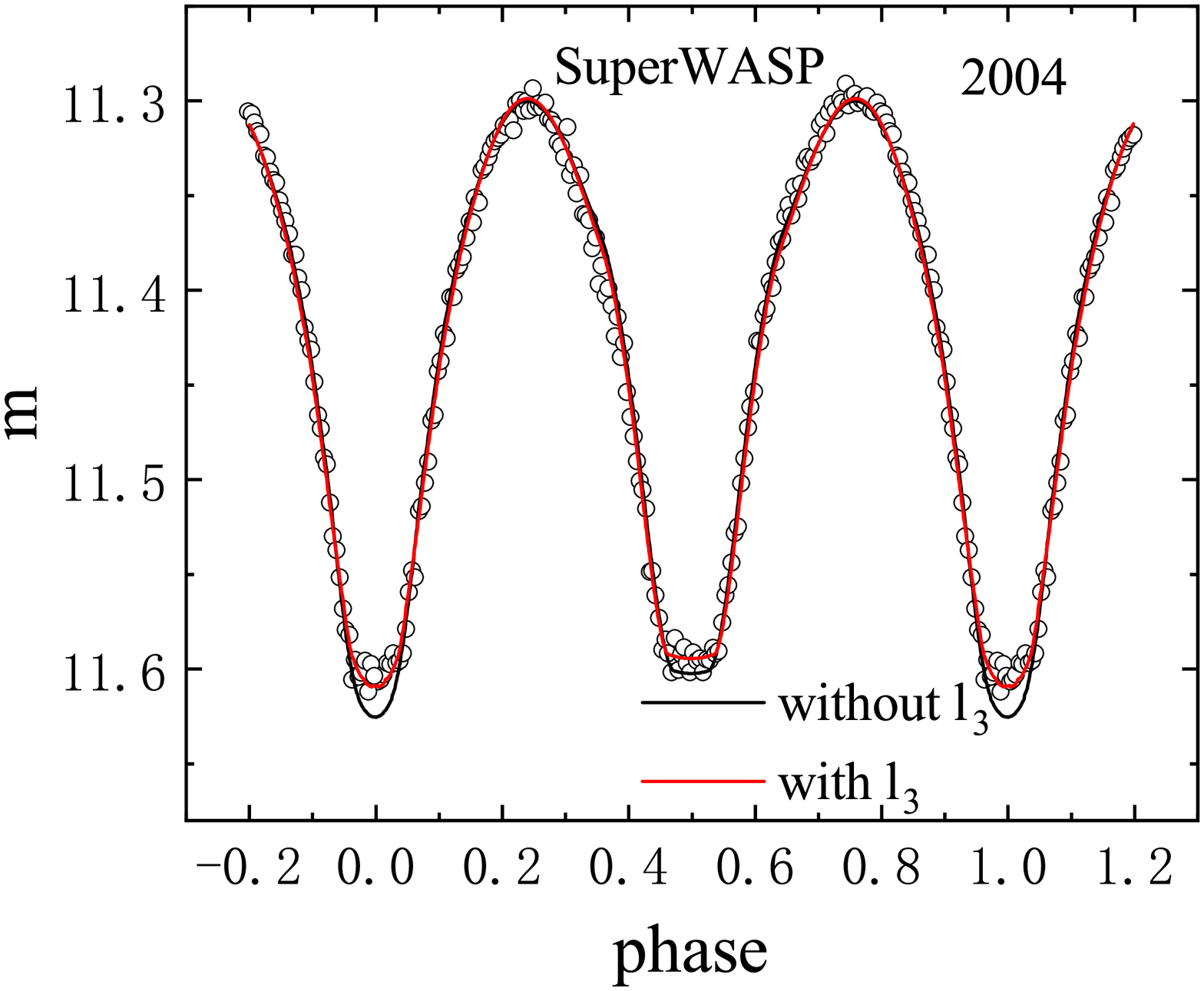}
\hspace{-1.1cm}
\includegraphics[width=0.36\textwidth]{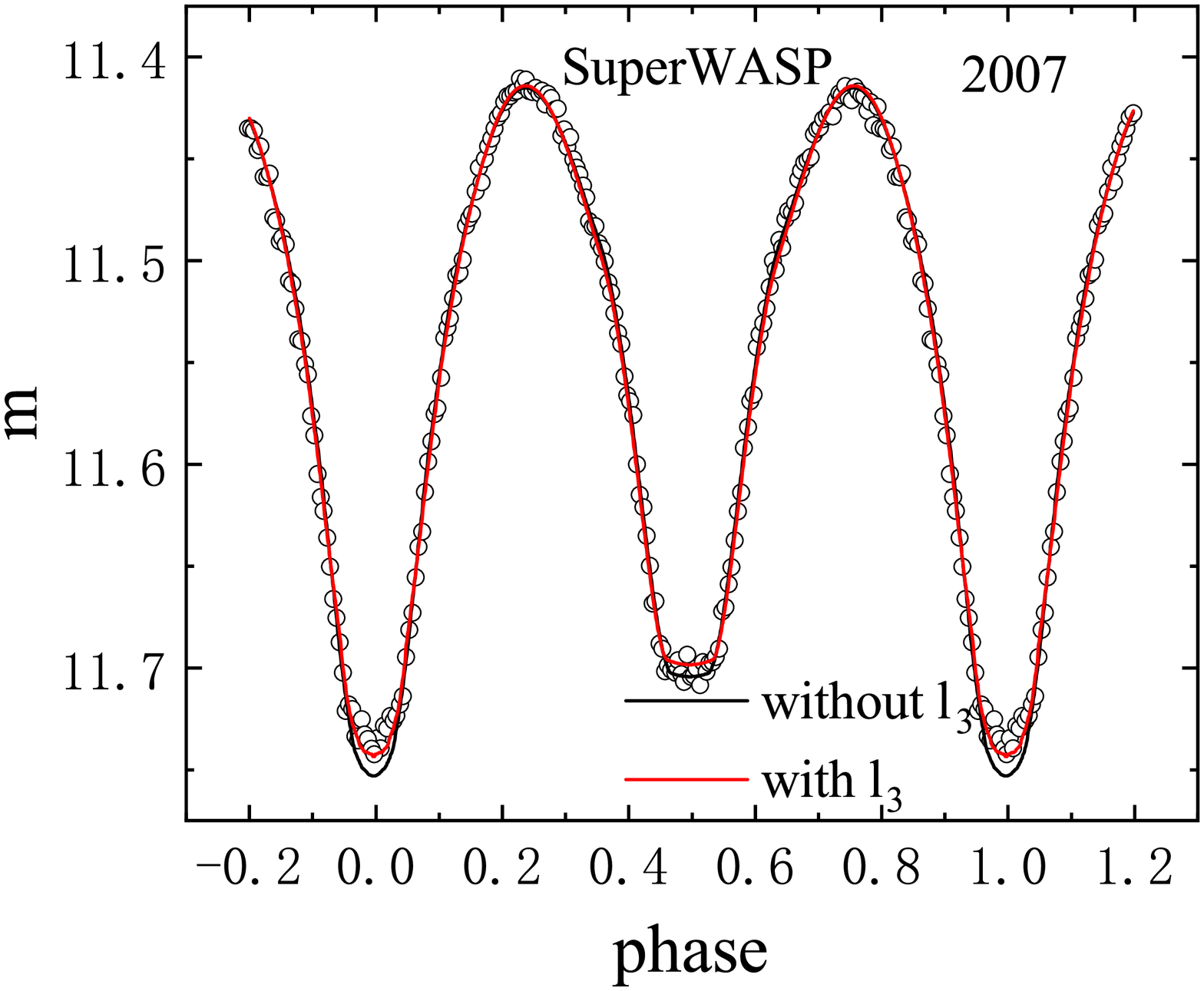}
\includegraphics[width=0.36\textwidth]{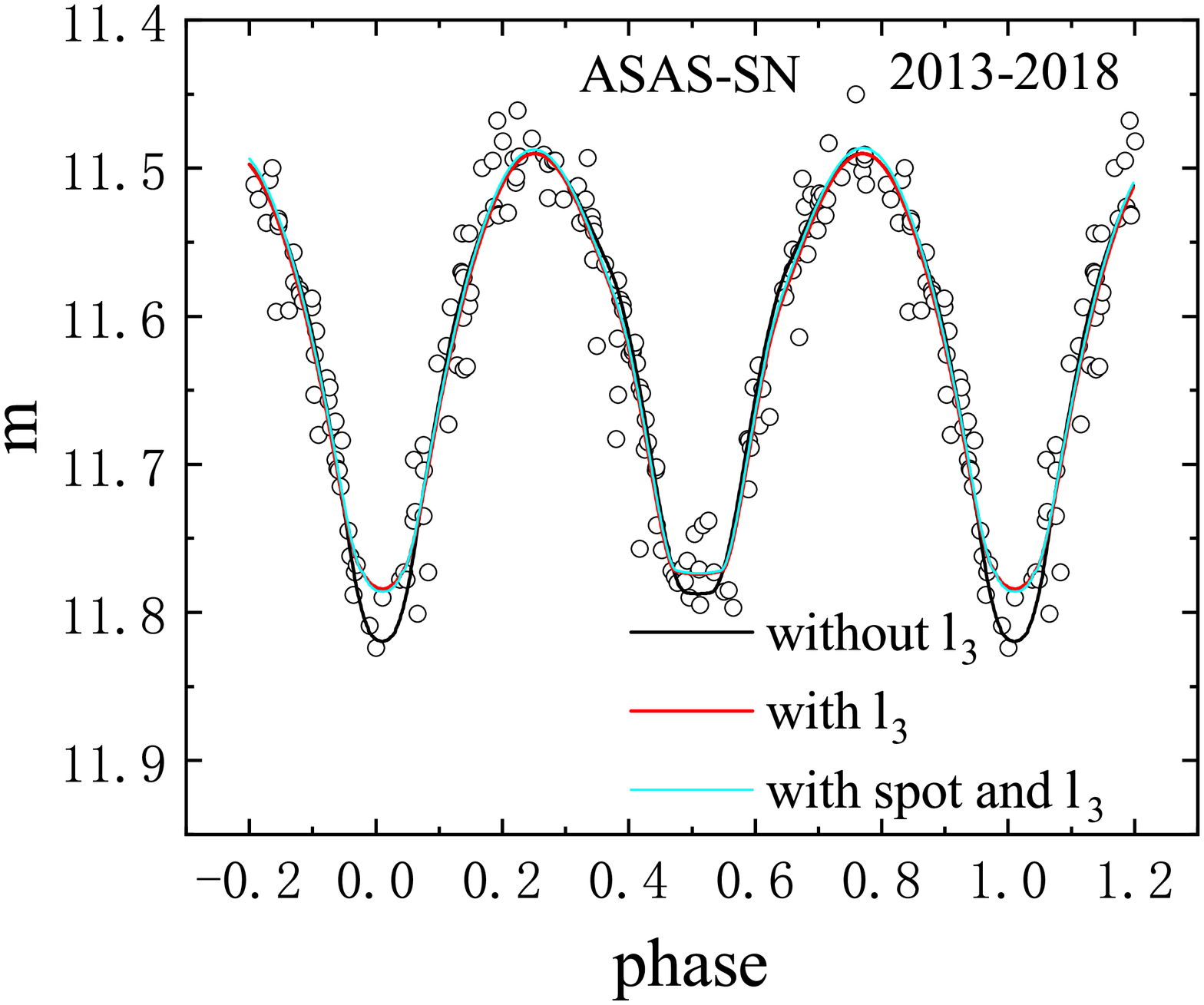}
\hspace{-1.1cm}
\includegraphics[width=0.36\textwidth]{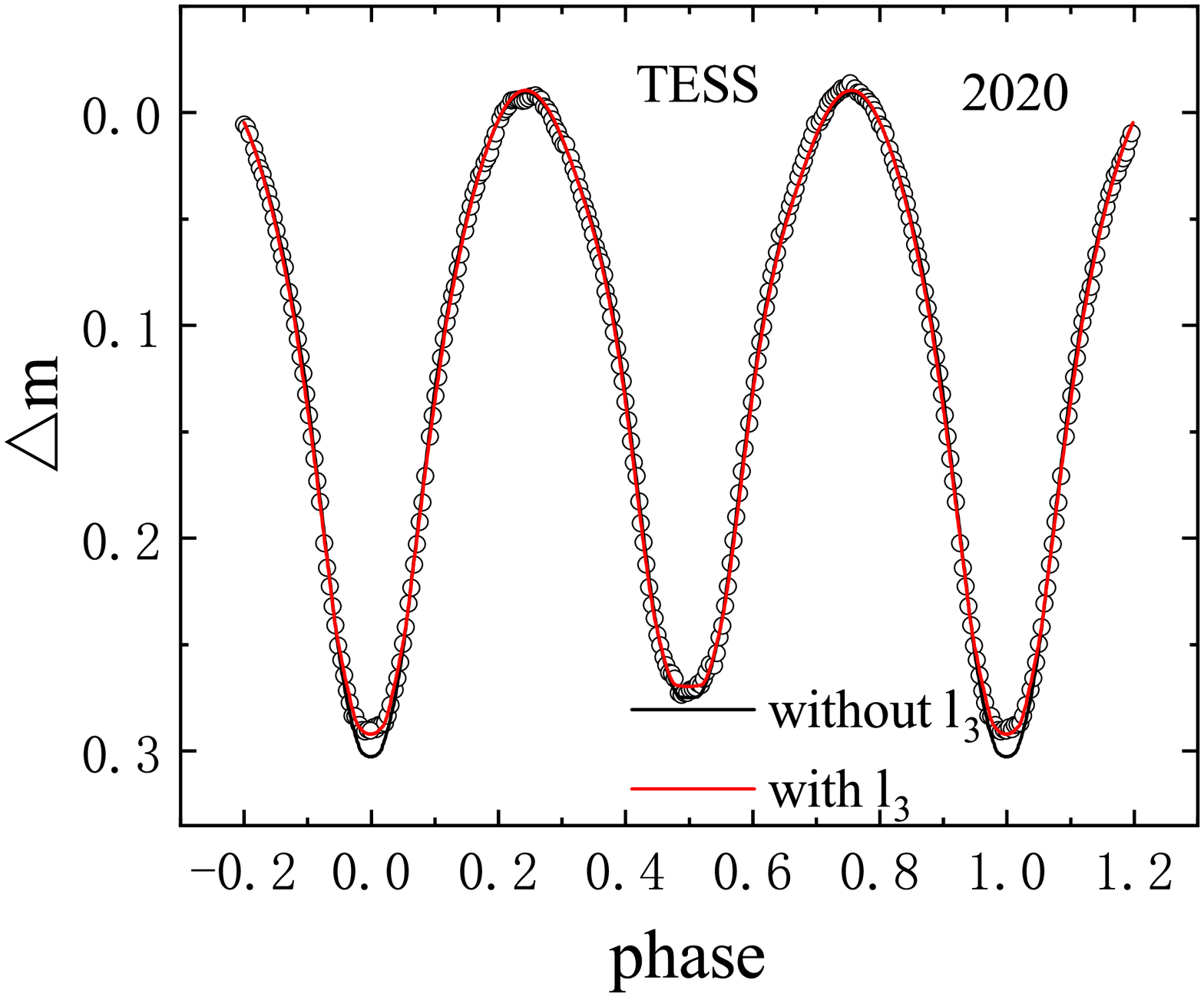}
\hspace{-1.1cm}
\includegraphics[width=0.36\textwidth]{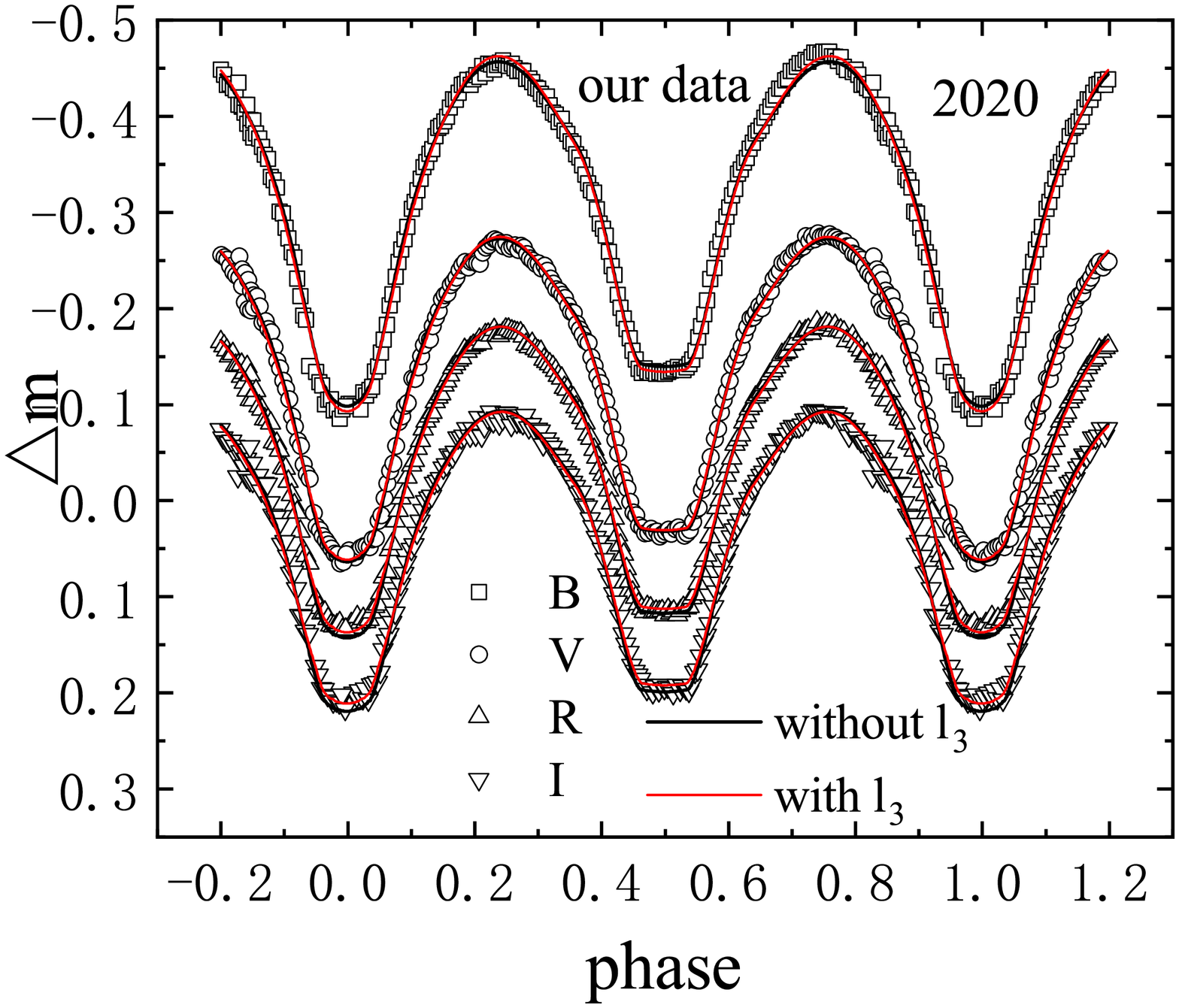}
\caption{Theoretical and observed light curves of NSVS 5029961 obtained by NSVS, SuperWASP, ASAS-SN, TESS and us. The observation times are also presented.}
\end{center}
\end{figure}

\renewcommand\arraystretch{1.3}
\begin{table}
\tiny
\caption{The Spectroscopic information of NSVS 5029961}
\begin{tabular}{p{1.7cm}p{2cm}p{1.2cm}p{1.5cm}p{1.3cm}p{1.4cm}p{1.6cm}p{2.1cm}p{2.1cm}}
\hline
Data        & HJD           & Phase   & Exposures & Subclass & $T_{eff}$ & log(g) & Radial velocity \\
(d)         & (d)           &         & (s)       &          & (K)       &        & $(km s^{-1})$   \\
\hline
2013 Mar 09 & 2456361.04682 & 0.24507 & 1200      & F6       & 6087.71   & 4.143  & -31.67           \\
2014 Apr 27 & 2456775.04536 & 0.84339 & 1800      & F6       & 6099.91   & 4.159  & -29.29           \\
2015 Jan 08 & 2457031.04466 & 0.11541 & 1800      & F5       & 6130.13   & 4.073  & -46.30           \\
2015 Feb 05 & 2457059.04616 & 0.49818 & 1800      & F6       & 6144.29   & 4.183  & -40.32           \\
2017 Mar 15 & 2457827.57197 & 0.68642 & 1800      & F9       & -         & -      & -                \\
2017 Apr 13 & 2457856.57123 & 0.56320 & 1800      & G2       & -         & -      & -                \\
\hline
\end{tabular}
\end{table}

\begin{figure*}
\begin{center}
\includegraphics[width=0.6\textwidth]{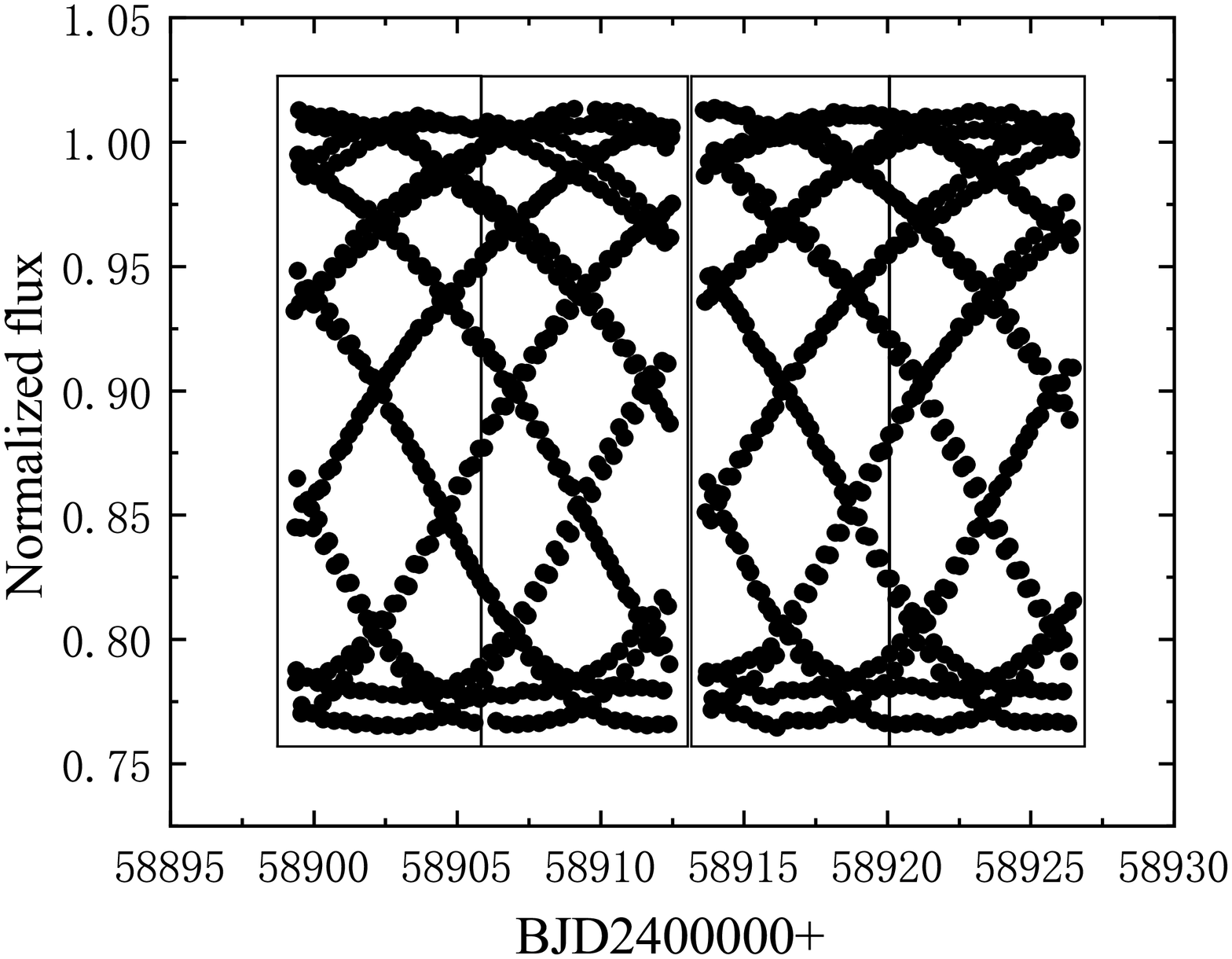}\\
\hspace{-1.04cm}
\includegraphics[width=0.3\textwidth]{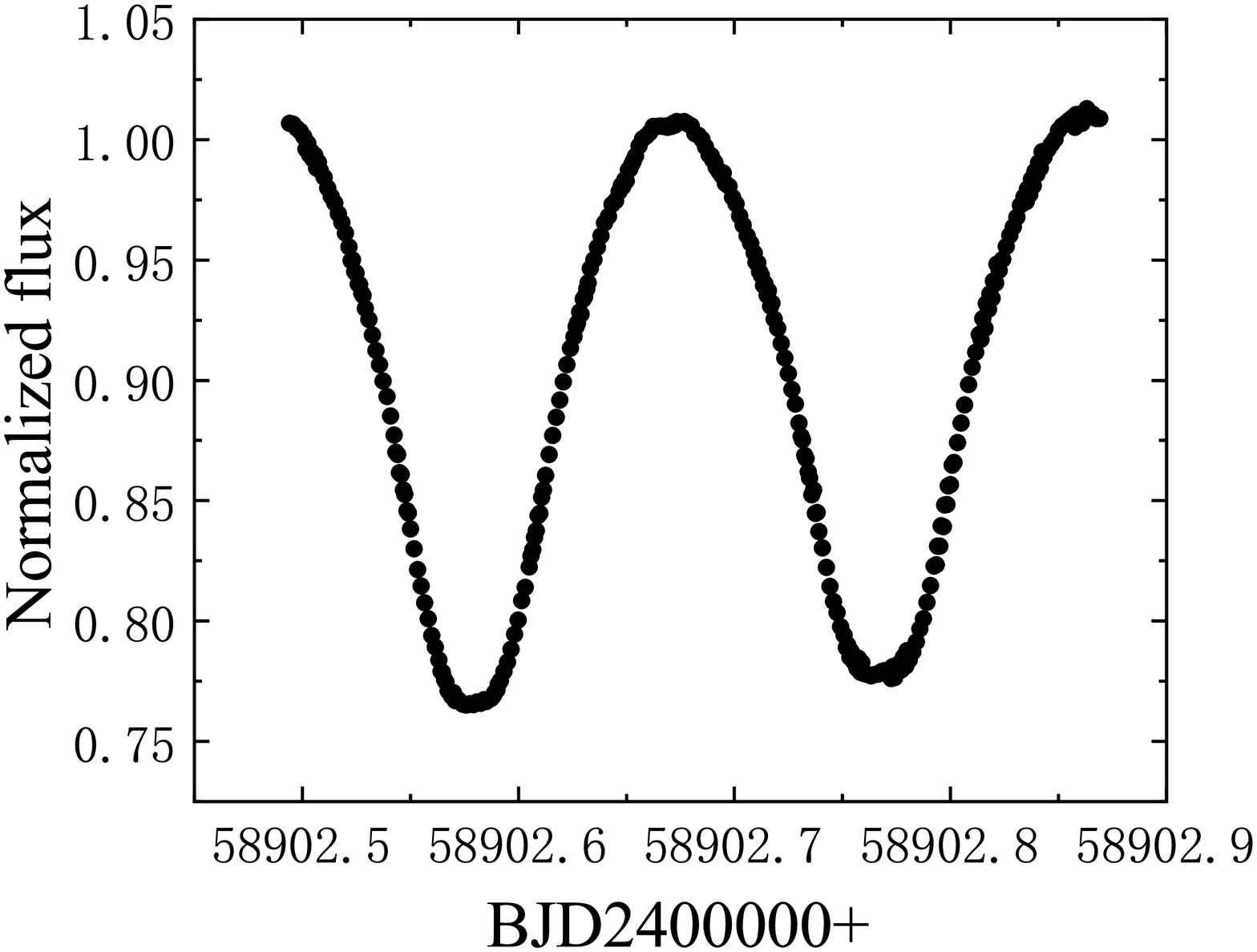}
\hspace{-1.04cm}
\includegraphics[width=0.3\textwidth]{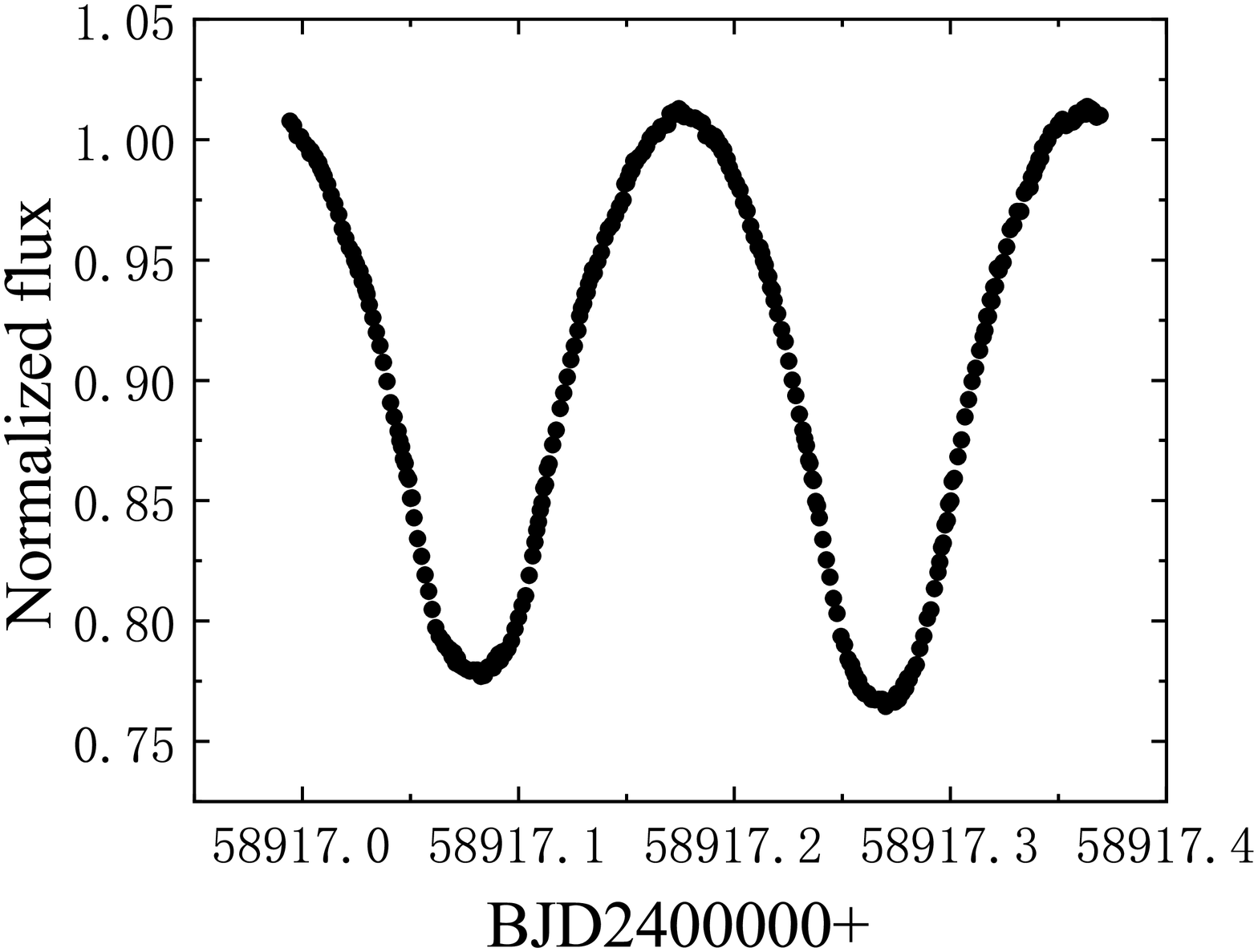}
\hspace{-1.04cm}
\includegraphics[width=0.3\textwidth]{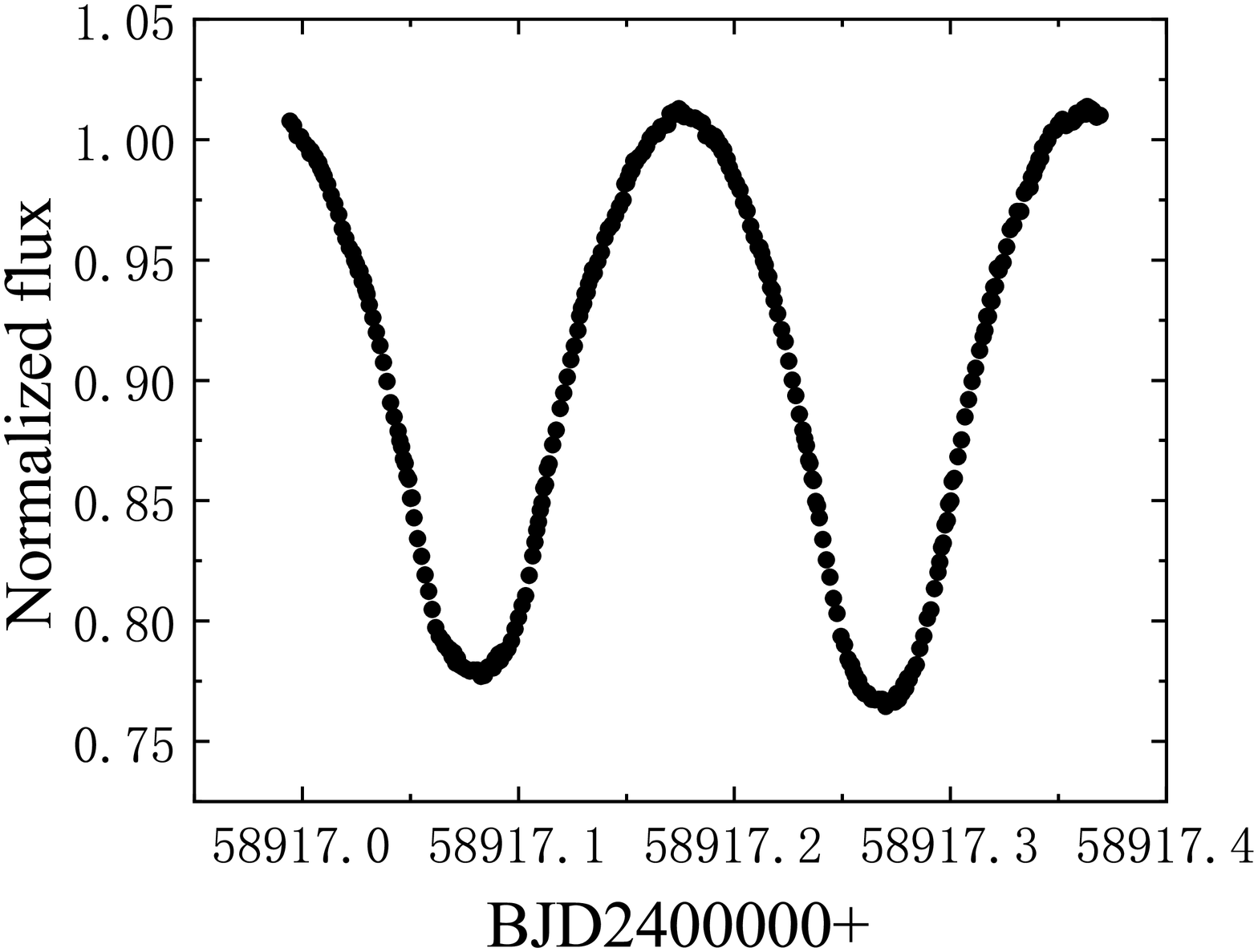}
\hspace{-1.04cm}
\includegraphics[width=0.3\textwidth]{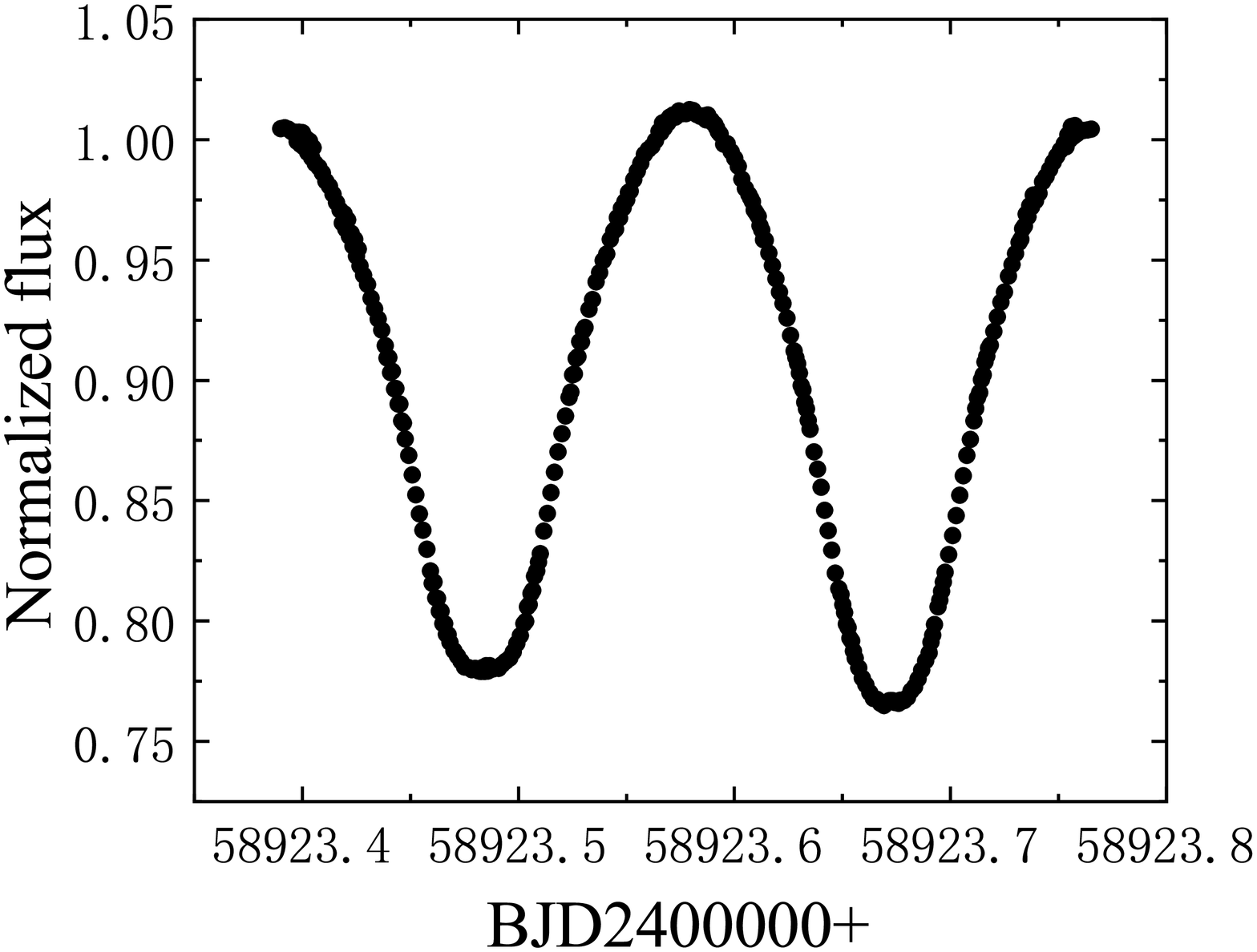}

\caption{The light curve of TESS (above) and the four parts transformed into one period (below).}
\end{center}
\end{figure*}

\renewcommand\arraystretch{1.2}
\begin{table*}
\tiny
\begin{center}
\caption{The minima timings of NSVS 5029961}
\begin{tabular}{p{1.4cm}p{0.9cm}p{1cm}p{1cm}p{0.95cm}p{0.45cm}|p{1.4cm}p{0.9cm}p{1cm}p{1cm}p{0.95cm}p{0.45cm}}
\hline
BJD       &Error &E   &O-C  &Residual  &Ref.  &BJD      &Error  &E &O-C  &Residual &Ref. \\
2400000+  &      &    &     &          &      & 2400000+ &        &   &      &          &      \\
\hline
53128.48365&0.00051&-15386.5&-0.04647&0.00177&(1)&	54171.65689&0.00038&-12617.0&-0.04030&-0.00073&(1)\\
53130.55199&0.00055&-15381.0&-0.04977&-0.00156&(1)&	54189.54808&0.00086&-12569.5&-0.04059&-0.00116&(1)	\\
53132.43642&0.00046&-15376.0&-0.04866&-0.00045&(1)&	54190.48945&0.00063&-12567.0&-0.04087&-0.00146&(1)	\\
53137.52229&0.00031&-15362.5&-0.04773&0.00043&(1)&	54194.63447&0.00065&-12556.0&-0.03914&0.00024&(1)	\\
53138.46395&0.00028&-15360.0&-0.04773&0.00042&(1)&	54195.57772&0.00062&-12553.5&-0.03755&0.00183&(1)	\\
53141.47752&0.00029&-15352.0&-0.04746&0.00067&(1)&	54202.54528&0.00037&-12535.0&-0.03825&0.00107&(1)	\\
53153.52992&0.00043&-15320.0&-0.04826&-0.00023&(1)&	54203.48841&0.00074&-12532.5&-0.03677&0.00254&(1)	\\
53154.47036&0.00038&-15317.5&-0.04948&-0.00146&(1)&	54204.42841&0.00028&-12530.0&-0.03843&0.00087&(1)	\\
53155.41289&0.00047&-15315.0&-0.04861&-0.00059&(1)&	54206.50060&0.00049&-12524.5&-0.03788&0.00140&(1)	\\
53157.48388&0.00082&-15309.5&-0.04926&-0.00127&(1)&	54208.56873&0.00070&-12519.0&-0.04140&-0.00213&(1)	\\
53158.42628&0.00069&-15307.0&-0.04852&-0.00053&(1)&	54210.45356&0.00034&-12514.0&-0.03988&-0.00063&(1)\\
53160.49651&0.00060&-15301.5&-0.04993&-0.00196&(1)&	54211.39774&0.00048&-12511.5&-0.03736&0.00189&(1)\\
53161.44222&0.00062&-15299.0&-0.04588&0.00208&(1)&	54212.52798&0.00040&-12508.5&-0.03711&0.00213&(1)	\\
53167.46678&0.00049&-15283.0&-0.04792&-0.00001&(1)&	54213.46702&0.00101&-12506.0&-0.03972&-0.00050&(1)	\\
53170.48115&0.00059&-15275.0&-0.04685&0.00104&(1)&	54214.40994&0.00049&-12503.5&-0.03846&0.00076&(1)	\\
53173.49308&0.00077&-15267.0&-0.04822&-0.00036&(1)&	54214.59780&0.00053&-12503.0&-0.03893&0.00029&(1)	\\
53174.43394&0.00047&-15264.5&-0.04902&-0.00116&(1)&	54215.53964&0.00065&-12500.5&-0.03875&0.00046&(1)	\\
53190.44294&0.00080&-15222.0&-0.04818&-0.00046&(1)&	54216.48037&0.00045&-12498.0&-0.03967&-0.00047&(1)	\\
53197.41143&0.00081&-15203.5&-0.04795&-0.00028&(1)&	54217.42193&0.00072&-12495.5&-0.03977&-0.00058&(1)	\\
54115.72022&0.00140&-12765.5&-0.04258&-0.00254&(1)&	54219.49297&0.00048&-12490.0&-0.04037&-0.00120&(1)	\\
54122.69098&0.00041&-12747.0&-0.04008&-0.00010&(1)&	54220.43590&0.00077&-12487.5&-0.03910&0.00007&(1)	\\
54135.68551&0.00076&-12712.5&-0.04041&-0.00053&(1)&	54223.45182&0.00056&-12479.5&-0.03648&0.00266&(1)	\\
54139.63923&0.00081&-12702.0&-0.04164&-0.00180&(1)&	54225.52046&0.00050&-12474.0&-0.03949&-0.00036&(1)	\\
54141.71366&0.00043&-12696.5&-0.03886&0.00096&(1)&	54226.46071&0.00065&-12471.5&-0.04089&-0.00177&(1)	\\
54142.65301&0.00034&-12694.0&-0.04117&-0.00135&(1)&	54227.40458&0.00049&-12469.0&-0.03868&0.00043&(1)	\\
54143.59476&0.00068&-12691.5&-0.04107&-0.00127&(1)&	54230.41755&0.00035&-12461.0&-0.03901&0.00008&(1)	\\
54145.67010&0.00298&-12686.0&-0.03738&0.00241&(1)&	54231.54753&0.00046&-12458.0&-0.03902&0.00006&(1)	\\
54146.60888&0.00035&-12683.5&-0.04025&-0.00047&(1)&	54232.48978&0.00038&-12455.5&-0.03842&0.00065&(1)	\\
54149.62282&0.00068&-12675.5&-0.03961&0.00014&(1)&	54233.43094&0.00045&-12453.0&-0.03892&0.00014&(1)	\\
54153.57823&0.00037&-12665.0&-0.03916&0.00056&(1)&	54235.50245&0.00034&-12447.5&-0.03905&-0.00001&(1)	\\
54153.76451&0.00032&-12664.5&-0.04121&-0.00149&(1)&	54236.44469&0.00031&-12445.0&-0.03847&0.00057&(1)	\\
54154.70834&0.00036&-12662.0&-0.03904&0.00068&(1)&	54249.43868&0.00077&-12410.5&-0.03934&-0.00041&(1)\\
54155.65072&0.00041&-12659.5&-0.03832&0.00139&(1)&	54251.51069&0.00097&-12405.0&-0.03897&-0.00006&(1)	\\
54156.59122&0.00063&-12657.0&-0.03947&0.00023&(1)&	54575.44443&0.00048&-11545.0&-0.03507&0.00115&(1)	\\
54157.72089&0.00072&-12654.0&-0.03979&-0.00010&(1)&	58902.57904&0.00006&-57.0&-0.00041&-0.00012&(2)	\\
54158.66315&0.00054&-12651.5&-0.03919&0.00050&(1)&	58902.76747&0.00009&-56.5&-0.00031&-0.00002&(2)	\\
54159.60349&0.00045&-12649.0&-0.04050&-0.00083&(1)&	58909.17116&0.00010&-39.5&0.00011&0.00035&(2)	\\
54160.54625&0.00084&-12646.5&-0.03940&0.00027&(1)&	58909.35866&0.00007&-39.0&-0.00072&-0.00048&(2)	\\
54162.61576&0.00079&-12641.0&-0.04153&-0.00188&(1)&	58917.08071&0.00010&-18.5&-0.00025&-0.00008&(2)	\\
54163.56060&0.00044&-12638.5&-0.03835&0.00129&(1)&	58917.26873&0.00006&-18.0&-0.00056&-0.00039&(2)	\\
54163.74757&0.00088&-12638.0&-0.03971&-0.00007&(1)&	58923.48379&0.00010&-1.5&-0.00044&-0.00032&(2)	\\
54165.63079&0.00059&-12633.0&-0.03980&-0.00018&(1)&	58923.67232&0.00007&-1.0&-0.00024&-0.00012&(2)	\\
54166.57419&0.00071&-12630.5&-0.03806&0.00156&(1)&	58924.04922&0.00021&0.0&0.00000&0.00011&(3)	\\
54167.70322&0.00060&-12627.5&-0.03902&0.00059&(1)&	58924.23749&0.00021&0.5&-0.00006&0.00005&(3)	\\
54169.58513&0.00085&-12622.5&-0.04042&-0.00083&(1)&				&	&    &     &   &  	\\
\hline
\end{tabular}
\end{center}
(1) SuperWASP; (2) TESS; (3) This paper.
\end{table*}

\begin{figure}
\begin{center}
\includegraphics[width=0.6\textwidth]{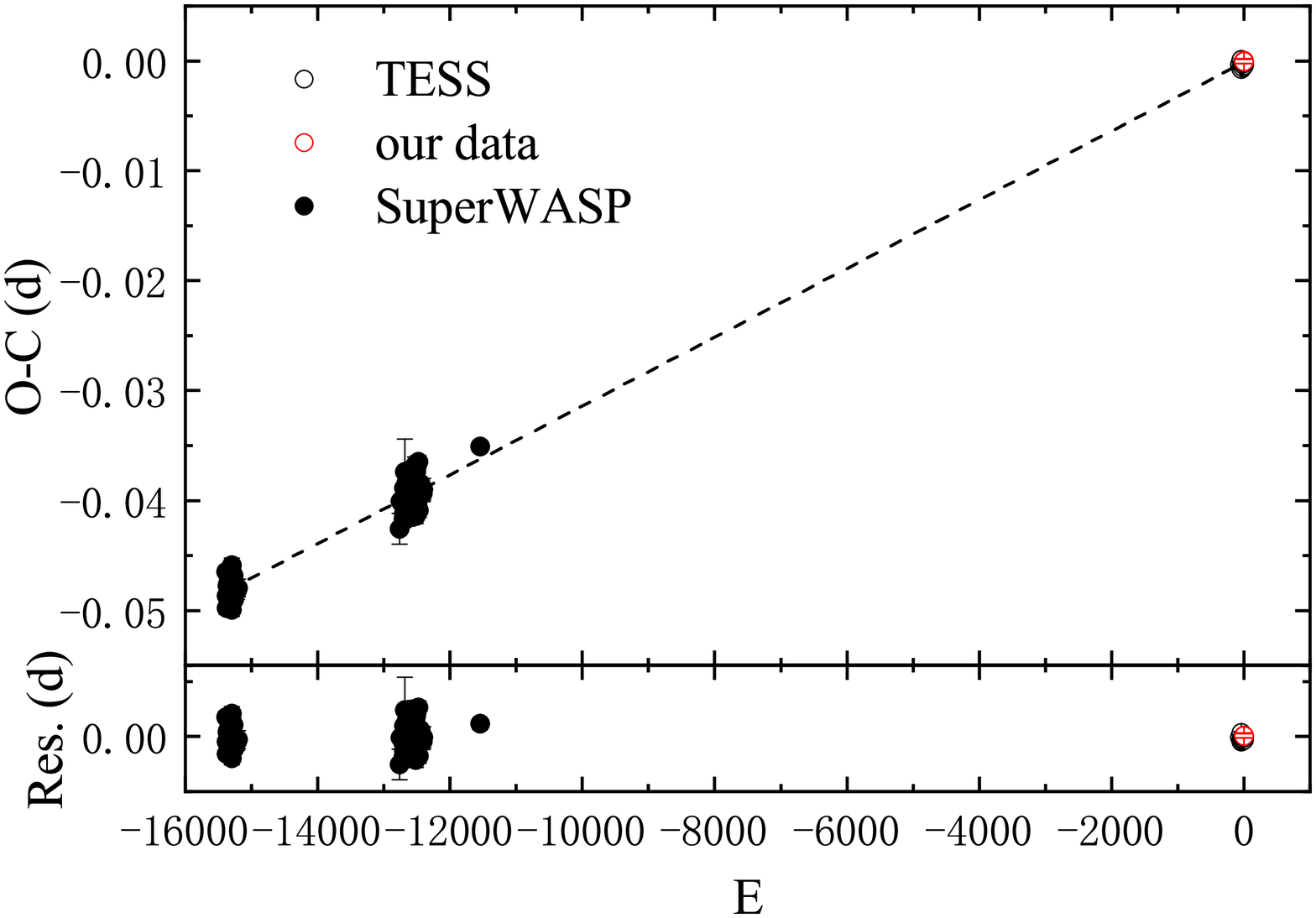}
\caption{The upper panel presents O-C diagram of the target. The lower panel displays the residuals.}
\end{center}
\end{figure}

\section{Photometric Solutions with W-D Method}
To determine the photometric solutions, we employed the 2013 version of Wilson-Devinney (W-D) code (Wilson \& Devinney 1971; Wilson 1979, 1990, 1994) to analyze the six sets of light curves. The effective temperature of primary star was estimated by using the color indices and LAMOST data. The color indices B - V = 0.530 mag, g - i = 0.299 mag, J - K = 0.291 mag were calculated based on the magnitudes of B, V, g and i bands from AAVSO Photometric All Sky Survey (APASS, Henden et al. 2016), and J and K bands from Two Micron All-Sky Survey (2MASS, Cutri et al. 2003). On account of interstellar extinction and reddening, we used E(B - V) (0.014 mag), E(g - i) (0.027 mag) and E(J - K) (0.007mag) determined from IRAS database\nolinebreak\footnotemark[5]\footnotetext[5]{https://irsa.ipac.caltech.edu/applications/DUST/} to calculate the dereddening indices. Then (B - V)$_0$ = 0.516 mag, (g - i)$_0$ = 0.272 mag and (J - K)$_0$ = 0.284 mag were derived. According to the color indices, Table 3 of Covey et al. (2007) and Table 5\nolinebreak\footnotemark[6]\footnotetext[6]{http://www.pas.rochester.edu/\url{~}emamajek/EEM\url{_}dwarf\url{_}UBVIJHK\url{_}colors\url{_}Teff.txt} of Pecaut \& Mamajek (2013), we obtained corresponding three temperatures (6240 K, 6510 K and 6170 K). We also calculated the average value of the four temperatures obtained by LAMOST which are listed in Table 3. Then we averaged the temperatures from dereddening color indices and LAMOST, and determined the temperature of the binary to be 6260$\pm170$ K, which was rounded off to three significant figures. It indicates a common convective envelope for the system, therefore the values of gravity darkening and bolometric albedo coefficients were taken as g$_{1,2}$ = 0.32 (Lucy 1967) and A$_{1,2}$ = 0.5 (Rucinski 1969) for both components. Based on Van Hamme's table (Van Hamme 1993), the logarithmic law was used to determine bolometric and bandpass limb-darkening coefficients. Mode 3 was used based on the contact configuration of the star. The adjustable parameters were the orbital inclination (i), the effective temperature of star 2 (T$_2$), the monochromatic luminosity of star 1 (L$_1$) and the dimensionless potential ($\Omega_1$ = $\Omega_2$).

Without radial-velocity measurements, the q-search method was carried out on our data to find the mass ratio and the result is shown in Figure 5. The minimum residual was found at q = 0.15. We set q as an adjustable parameter and chose it as the initial value. The better convergent solution was obtained finally and listed in Table 5. The corresponding synthetic light curves are presented in Figure 2. We also set l$_3$ as an adjustable parameter to search for the third light, but the third light was always negative, so the convergent solution cannot be obtained. The five sets of light curves obtained by TESS, SuperWASP, ASAS-SN, NSVS were analysed based on the result above. The mass ratio was set as a fixed parameter. Specially, because of 30-minute observation cadence, we took into account phase smearing effect for TESS data. We found that the theoretical curves of five sets did not fit the observed data well, so we added the third light. Better theoretical light curves were obtained from Figure 2 and the photometric solutions with l$_3$ are also displayed in Table 5. From Table 5, we found the degree of contact for our new light curves is smaller than that for all the others. Then, we tried to find the correlations between some specific parameters. The diagrams of $L_3/L - i$ and $L_3/L - f$ for all the light curves are shown in Figure 6. The correlation between inclination ($i$) and the proportion of the third luminosity in the total luminosity ($L_3/L$) is not obvious, while the degree of contact ($f$) is positively associated with $L_3/L$. The bigger the $L_3/L$, the greater the degree of contact ($f$). Therefore, the smallest degree of contact of our new light curves may be caused by the non-existent third light. The light curves of both TESS and our new observations were obtained in 2020, so the no third light solution of the latter is puzzling. Trying to explain it, we managed to start new solutions with positive third light of our new light curves by keeping the mass ratio and potential, $\Omega$, fixed as their TESS values. A convergent solution with positive third light was obtained successfully, and the determined parameters are listed in Table 5. The results of the adjustable parameters are similar to the previous values with no third light, and the amount of third light is very close to that of TESS solution. The fitted light curves with third light are displayed in Figure 2, and we can see that the fitted light curves with or without third light are nearly overlapped.

\begin{figure}\centering
\includegraphics[width=0.6\textwidth]{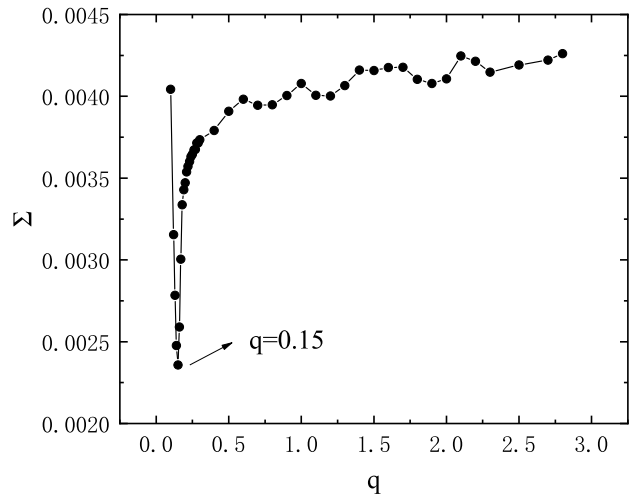}
\caption{Relationship between $\Sigma W_i(O-C)^2_i$ and the mass ratio q.}
\end{figure}

\renewcommand\arraystretch{1.3}
\begin{table*}
\tiny
\caption{Photometric elements of the target}
\begin{center}
\begin{tabular}{p{1.4cm}p{1.7cm}p{1.7cm}p{1.7cm}p{1.9cm}p{1.9cm}p{1.7cm}p{1.7cm}}
\hline
Parameters    &\multicolumn{2}{c}{Our data}           &TESS          &SuperWASP2004 &SuperWASP2007 &ASAS-SN       &NSVS           \\
              &(without l$_3$)              &(with l$_3$)        &              &              &              &              &    \\
\hline
$T_1(K)$      &6260                         &6260                &6260          &6260          &6260          &6260          &6260           \\
$T_2(K)$      &6192(6)                      &6206(6)             &6174(4)       &6314(17)      &6113(13)      &6446(134)     &6385(80)       \\
$q(M_2/M_1)$  &0.1515(0.0006)               &0.1515              &0.1515        &0.1515        &0.1515        &0.1515        &0.1515         \\
i             &78.0(0.2)                    &77.3(0.2)           &77.9(0.1)     &78.1(0.3)     &77.9(0.3)     &78.6(2.1)     &77.1(1.5)  \\
$\Omega$      &2.089(0.003)                 &2.074               &2.074(0.001)  &2.077(0.003)  &2.069(0.003)  &2.033(0.028)  &2.051(0.016)   \\
$L_1/L^a$     &-                            &-                   &0.7996(0.0047)&0.7574(0.0251)&0.7938(0.0217)&-             &0.7183(0.1217) \\
$L_1/L(B)^a$  &0.8512(0.0018)               &0.8439(0.0100)      &-             &-             &-             &-             &-              \\
$L_1/L(V)^a$  &0.8492(0.0016)               &0.8211(0.0102)      &-             &-             &-             &0.6216(0.1645)&-              \\
$L_1/L(R_c)^a$&0.8483(0.0015)               &0.8079(0.0107)      &-             &-             &-             &-             &-              \\
$L_1/L(I_c)^a$&0.8475(0.0014)               &0.7988(0.0113)      &-             &-             &-             &-             &-           \\
$L_3/L^a$     &-                            &-                   &0.0547(0.0020)&0.0929(0.0116)&0.0691(0.0094)&-             &0.1260(0.0618)  \\
$L_3/L(B)^a$  &-                            &0.0031(0.0038)      &-             &-             &-             &-             &-            \\
$L_3/L(V)^a$  &-                            &0.0282(0.0042)      &-             &-             &-             &0.2329(0.1043)&-             \\
$L_3/L(R_c)^a$&-                            &0.0431(0.0047)      &-             &-             &-             &-             &-             \\
$L_3/L(I_c)^a$&-                            &0.0531(0.0051)      &-             &-             &-             &-             &-           \\
$r_1$         &0.5530(0.0007)               &0.5584              &0.5584(0.0003)&0.5577(0.0013)&0.5605(0.0011)&0.5737(0.0115)&0.5678(0.0063)   \\
$r_2$         &0.2389(0.0029)               &0.2449              &0.2449(0.0003)&0.2438(0.0016)&0.2474(0.0014)&0.2684(0.0197)&0.2569(0.0090)  \\
f             &19.1(2.6)$\%$                &33.8$\%$            &33.8(0.7)$\%$ &31.3(3.5)$\%$ &39.1(3.0)$\%$ &76.3(28.7)$\%$&57.4(16.3)$\%$ \\
\hline
\end{tabular}
\end{center}
$^a$ $L_1$ is the luminosity of star 1 obtained from the differential corrections program (DC) of Wilson-Devinney program, $L_3$ is the estimated third luminosity based on $L_3$ = 4$\pi \times $l$_3$, where $l_3$, the third light, is also obtained from DC. $L$ refers to the total luminosity of the system which is calculated by $L$ = $L_1$ + $L_2$ + $L_3$. $L_1/L$ and $L_3/L$ for TESS, SuperWASP and NSVS are in their corresponding band. Our data have $L_1/L$ and $L_3/L$ in BV(RI)$_c$ bands and ASAS-SN data have $L_1/L$ and $L_3/L$ only in V band.
\end{table*}

\begin{figure}\centering
\includegraphics[width=0.42\textwidth]{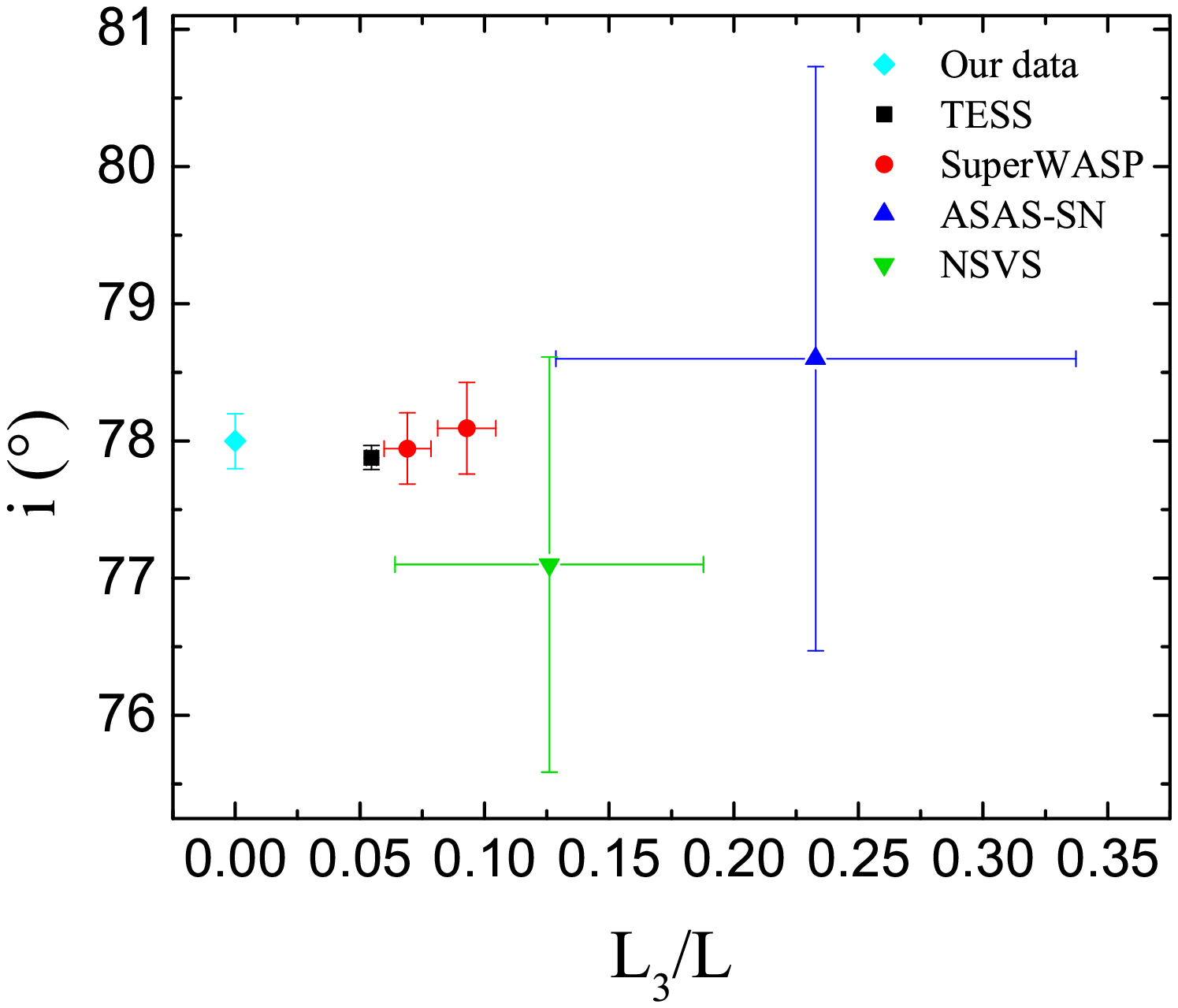}
\includegraphics[width=0.433\textwidth]{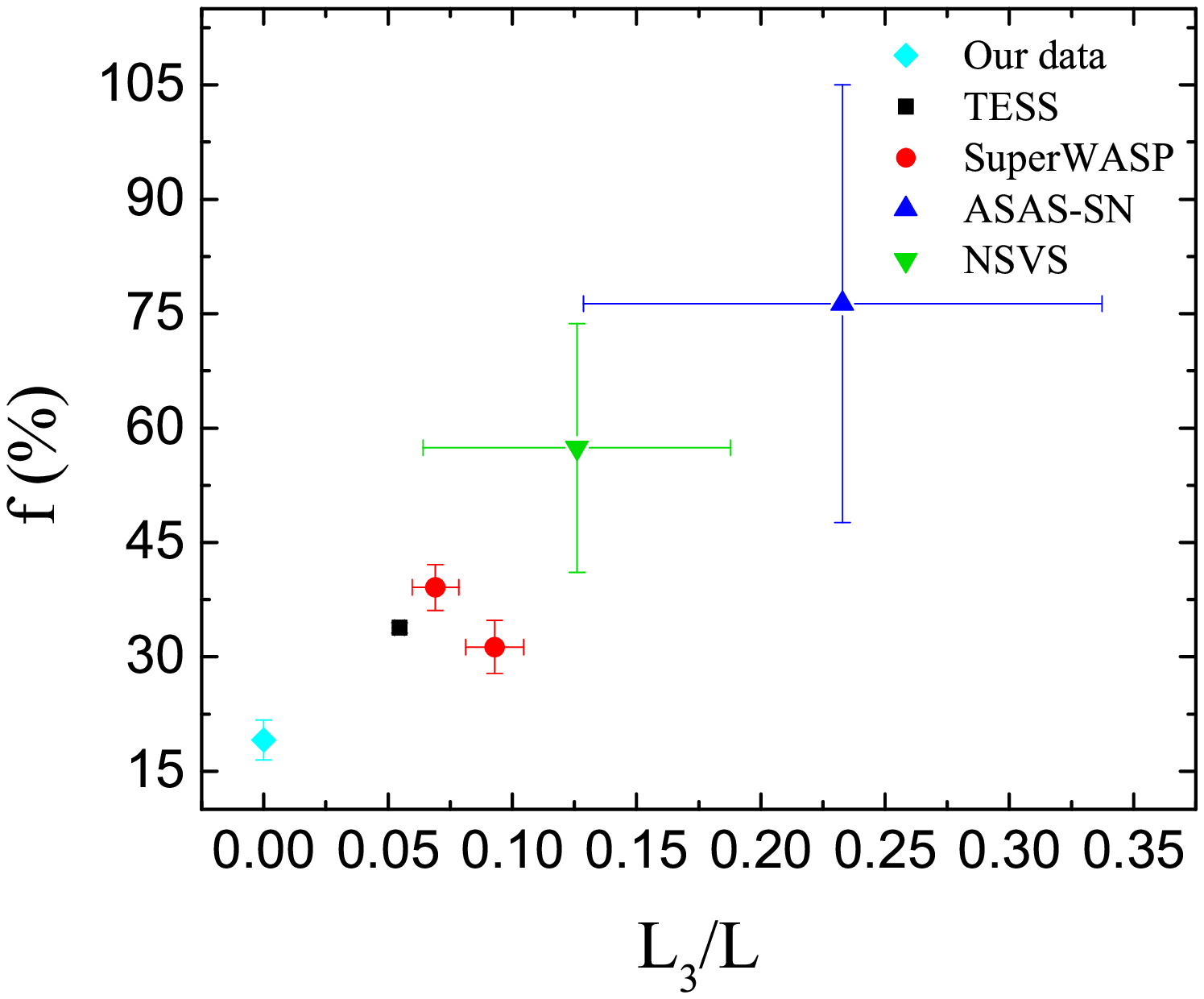}
\caption{The $L_3/L$ - i diagram (left) and the $L_3/L$ - f diagram (right). }
\end{figure}

\section{Spectral Analysis}
The excess emission lines H$_\alpha$, H$_\beta$, Ca II H$\&$K and IRT etc are applied as the optical and near-infrared diagnostics of stellar chromospheric activity in late type rotating stars (Wilson 1978; Barden 1984; Montes et al. 1995). In order to analyse these lines of the six spectra, the spectral substraction technique (Barden 1985) was used to remove photospheric lines from the observed spectra. We first tried to find the synthesized spectrum containing photospheric information. Some stars near the temperature of the primary star of NSVS 5029961 were selected from a catalogue of inactive radial velocity standard stars (Huang et al. 2018). The selected stars were searched in LAMOST DR 7 then and 11 template stars were determined. Since the downloaded object and template spectra had been reduced by the LAMOST data pipelines (Luo et al. 2012), we only needed to normalize them with the continuum package of IRAF\nolinebreak\footnotemark[7]\footnotetext[7]{IRAF is distributed by the National Optical Astronomy Observatories, which is operated by the Association of Universities for Research in Astronomy Inc., under contract to the National Science Foundation (http://iraf.noao.edu).}. According to the standard deviations obtained by subtracting the template spectra from the object spectra, the best template spectrum was determined to be 2MASS 10501536+4338441 with temperature 6487 K and spectral type F5, and set as the synthesized spectrum. The process of matching the template spectrum above was outside the H$_\alpha$ lines. Then the subtracted spectra were ultimately obtained by subtracting the inactive synthesized spectrum from the object spectra. Figure 7 plots the object, synthesized and subtracted spectra. It indicates that only four subtracted spectra have good enough SNR to study excess emission lines. We identified the He I, Ca II H$\&$K, H$_\delta$, H$_\gamma$, H$_\beta$, H$_\alpha$ and Ca II infrared triplet lines from the subtracted spectra of Figure 7. Because the red and blue ends of the spectra have low S/N, the equivalent widths (EWs) of only parts of these lines were calculated by using Splot package of IRAF and listed in Table 6. The excess emission lines indicate that the system has high chromospheric activity. Considering that the photometric light curve of ASAS-SN was obtained at similar observation time of LAMOST spectra, we adopted the spot model to fit it. A well fitted result was obtained by adding the third light and one spot on the secondary star. The theoretical curve and convergent solution are shown in Figure 2 (the cyan line) and Table 7, respectively. This is consistent with the result of spectral analysis. The presence of the spot also demonstrates the magnetic activity of the binary star.

\begin{figure}\centering
\includegraphics[width=0.52\textwidth]{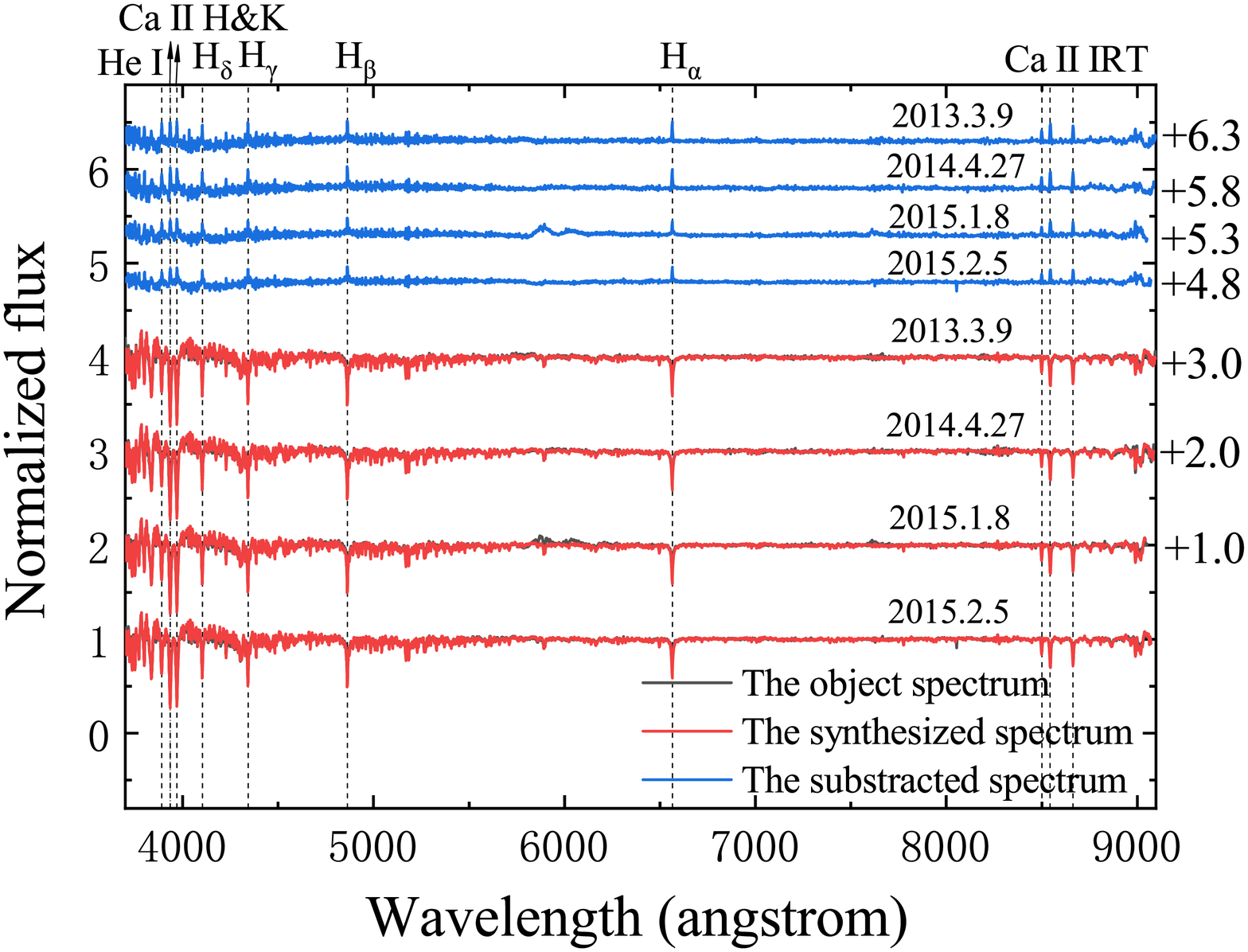}
\hspace{-1.5cm}
\includegraphics[width=0.52\textwidth]{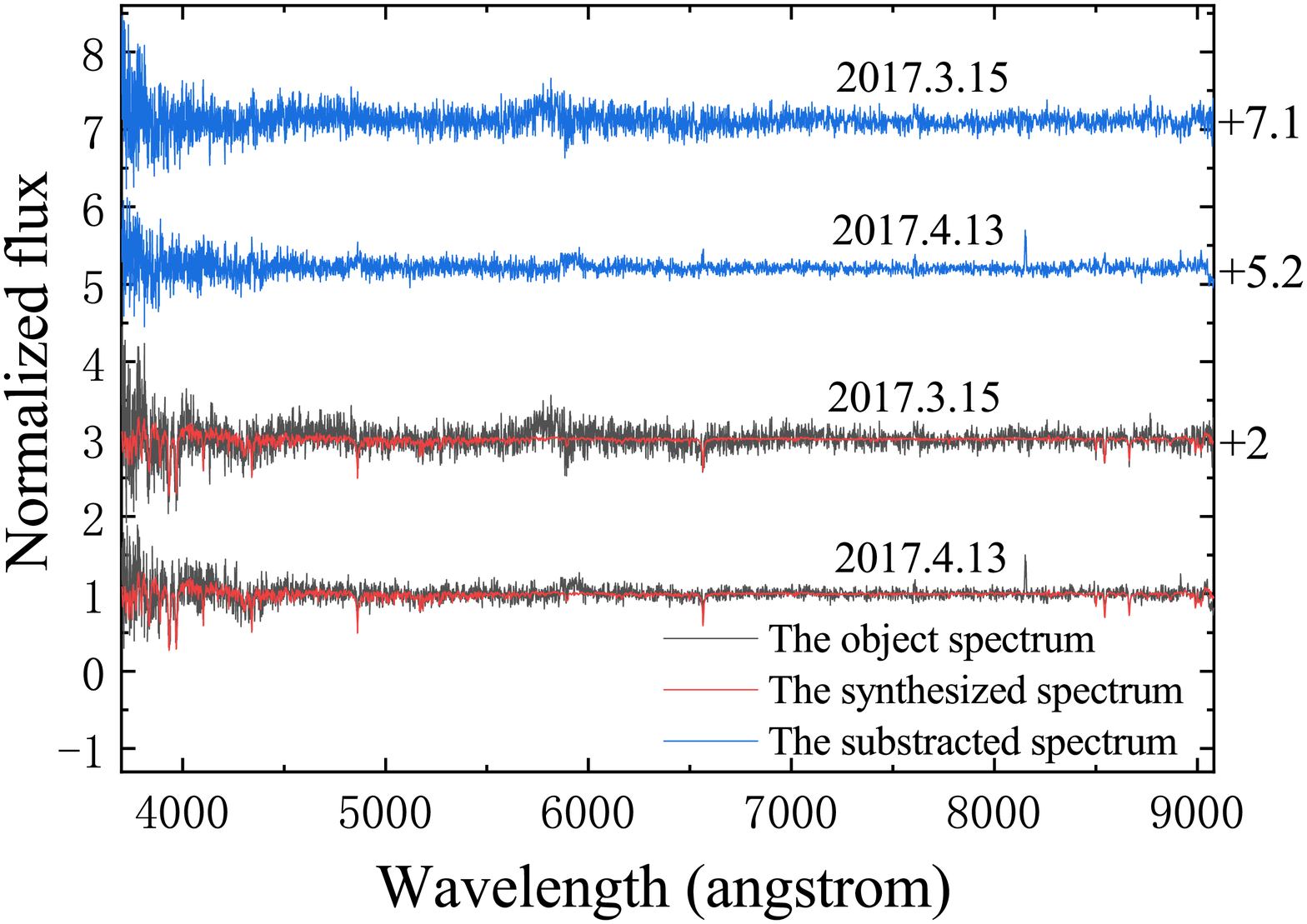}
\caption{The normalized object, synthesized and subtracted spectra observed by LAMOST. The excess emission lines are marked with dotted lines.}
\end{figure}

\renewcommand\arraystretch{1.3}
\begin{table}
\tiny
\caption{The EWs of chromospheric activity indicators}
\begin{tabular}{p{1.5cm}p{2cm}p{2cm}p{2cm}p{2cm}p{2cm}p{2cm}}
\hline
Data(d)      & H$\gamma$ (4341.68{\AA}) & H$\beta$ (4862.68{\AA}) & H$\alpha$ (6564.61{\AA}) & Ca II (8500.35{\AA}) & Ca II (8544.44{\AA}) & Ca II (8664.52{\AA})  \\
\hline
2013 Mar 09  & 3.454(0.009)     & 3.400(0.052)    & 4.646(0.034)     & 5.093(0.023)   & 5.579(0.009)   & 5.582(0.019)  \\
2014 Apr 27  & 4.334(0.014)     & 3.099(0.027)    & 5.073(0.019)     & 5.067(0.020)   & 5.723(0.009)   & 5.745(0.017)  \\
2015 Jan 08  & 3.591(0.051)     & 2.999(0.016)    & 4.315(0.036)     & 4.750(0.011)   & 6.154(0.092)   & 4.905(0.107)  \\
2015 Feb 05  & 4.078(0.025)     & 3.058(0.039)    & 4.491(0.172)     & 4.768(0.024)   & 5.100(0.016)   & 5.094(0.014)  \\
\hline
\end{tabular}
\end{table}

\begin{table}
\centering
\caption{The photometric solution with third light and one spot on the secondary star for the light curve of ASAS-SN}
\begin{tabular}{p{2.5cm}p{2cm}p{2cm}}
\hline
Parameters            & Values      &Error\\
\hline
$T_1(K)$              & 6260        &  -  \\
$T_2(K)$              & 6410        &133\\
$q(M_2/M_1)$          & 0.1515      &  - \\
i                     & 78.1        &2.0   \\
$\Omega$              & 2.037       &0.026  \\
$L_1/L$(V)            & 0.6470      &0.1677 \\
$L_3/L$               & 0.2065      &0.0999 \\
$r_1$                 & 0.5729      &0.0070  \\
$r_2$                 & 0.2654      &0.0174  \\
f($\%$)               & 71.7        &27.0  \\
Spot                  & Star 2      &  -       \\
$\theta$(rad)         & 2.194       &1.360      \\
$\lambda$(rad)        & 0.762       &2.180     \\
r$_s$(rad)            & 0.168       &0.283      \\
T$_s$                 & 0.806       &0.852      \\
\hline
\end{tabular}
\end{table}
\section{Discussions and Conclusions}
Although NSVS 5029961 had been monitored by many sky surveys, it was neglected for further study. The photometric and spectral investigation were conducted by us to understand its structure and evolution. We obtained the light curves in BV(RI)$_c$ bands of NSVS 5029961 and determined the photometric solutions by analyzing the light curves (including the sky survey data) with the W-D method. The parameters especially the degree of contact differ greatly among different surveys from Table 5. The result of our data is more accurate because it was obtained through the four-band photometry analysis with high precision, and we took it as the final result. The photometric solutions indicate that NSVS 5029961 is an A-subtype shallow contact binary with a contact degree of 19.1 $\pm$ 2.6\% and a mass ratio of 0.1515 $\pm$ 0.0006. The symmetry of the light curves and the sharp q-search curve bottom mean that the mass ratio has small uncertainty and the solutions are generally reliable (Pribulla et al. 2003; Terrell \& Wilson 2005; Zhang et al. 2017). We found that the degree of contact has a positive relationship with the proportion of the third luminosity in the total luminosity from Figure 6. For the photometric results of our new light curves, the degree of contact is the smallest, which may be due to the non-existence of the third light. The temperature difference between the two components (68 K) means that they are under thermal contact. The spectral type of the primary star is F7, while the one of the other is F8 according to Table 5\nolinebreak\footnotemark[6] of Pecaut \& Mamajek (2013). The presence of magnetic activity indicators in the subtracted spectra indicates that there is strong magnetic activity in the system, which is also supported by the well fit of ASAS-SN light curve with the spot model. The O-C diagram was analyzed by using all available times of minimum light. By using a linear fit, we corrected the period to be 0.37666613($\pm$0.00000003) days.

Without radial velocity curves for NSVS 5029961, we adopted the distance given by Gaia mission (Gaia Collaboration et al. 2018; Bailer-Jones et al. 2018) to estimate its absolute parameters. The maximum of V-band visual magnitude (m$_{Vmax}$) was estimated from the data observed by ASAS-SN. m$_{V1}$ - m$_{Vmax}$ = -2.5 log (L$_1/$L) and m$_{V2}$ - m$_{Vmax}$ = -2.5 log (L$_2/$L) were used to derive the V-band visual magnitudes of the two components. Applying the Gaia distance (D), m$_{V1,2}$ and the extinction coefficient (A$_V$) found from IRAS database\nolinebreak\footnotemark[5] to the equation, M$_{V1,2}$ = m$_{V1,2}$ - 5 log D + 5 - A$_V$, we determined the absolute magnitudes. The absolute bolometric magnitudes (M$_{bol1,2}$) were calculated based on M$_{bol1,2}$ = M$_{V1,2}$ + BC$_{V1,2}$, where the bolometric corrections (BC$_{V1,2}$) were obtained from Table 5\nolinebreak\footnotemark[6] of Pecaut \& Mamajek (2013). Then M$_{bol1,2}$ = -2.5 logL$_{1,2}$/L$_\odot$ + 4.74 was used to derive the luminosities of the two components. According to the Stefan-Boltzmann law (L = 4$\pi\sigma$T$^4$R$^2$), we derived the radii of the two components (when calculating the errors of the radii of the two components, the errors of T$_1$ and T$_2$ were considered, the error of T$_1$ is 170 K, and the error of T$_2$ was computed to be 174 K by the error transfer formula), which were then divided by the relative radii obtained from photometric solutions to get two values of semimajor axis. The average of the two values was computed as the final semimajor axis. The masses were determined with the mass ratio presented in Table 5 and the third law of Kepler, M$_1$ + M$_2$ = 0.0134a$^3$/P$^2$. All absolute parameters and errors on basis of error transfer formulas are shown in Table 8.

We collected the mass ratios, fill-out factors, temperatures of primary and secondary stars, absolute parameters and rates of periodic variation of the contact binaries with mass ratios below 0.25 and fill-out factors below 25\%. The information is listed in Table 9. The equation (Yang \& Qian 2015)
\begin{equation}
\frac{J_{spin}}{J_{orb}} = \frac{1+q}{q}[(k_1r_1)^2+(k_2r_2)^2q] \\
\end{equation}
was used to calculate the ratio of the spin angular momentum to the orbital angular momentum (J$_{spin}$/J$_{orb}$). The values of k$_{1,2}^2$ were set as 0.06 (Li \& Zhang 2006). J$_{spin}$/J$_{orb}$ $<$ 1/3 was determined for all targets, indicating they are in a stable state at present. To have a better understanding of the evolutionary status of NSVS 5029961, we present the mass-luminosity and mass-radius distributions of our target and the binaries mentioned above in Figure 8. The filled symbols denote the primary components and the open symbols denote the secondary components. The two components of our target are marked as red circles. The zero age main sequence (ZAMS) and terminal age main sequence (TAMS) lines (Hurley et al. 2002) are represented by solid and dashed lines, respectively. From Figure 8, we can see that the evolutionary states of the two components of all binaries are very similar. The more massive component lies near the ZAMS, meaning that it is a little evolved star, while the secondary component is located above the TAMS, implying that it has evolved away from the main sequence.

The initial masses of both components are of great importance to study the evolution of contact binaries, so we calculated them with the method developed by Y{\i}ld{\i}z \& Do\u{g}an (2013). The method is based on stellar models with mass loss and assumes that mass transfer and loss start near the TAMS phase of the secondary component. The initial mass of the secondary star (M$_{i2}$) was computed as 1.79 $\pm$ 0.24 M$_\odot$ in terms of the luminosity excess for its current mass by using the equation,
\begin{equation}
\ M_{i2} = M_2+2.50(M_L-M_2-0.07)^{0.64}, \\
\end{equation}
where M$_2$ is the present mass of the secondary component and M$_L$ was computed with M$_L$ = $(\frac{L_2}{1.49})^{\frac{1}{4.216}}$. Then from the physical constraint on the reciprocal of the initial mass ratio, the initial mass of the primary star (M$_{i1}$ = 1.36 $\pm$ 0.18 M$_\odot$) can also be determined. Y{\i}ld{\i}z \& Do\u{g}an (2013) found A-subtype binaries usually have the initial masses of the secondary stars between 1.7 and 2.6 M$_\odot$, while W-subtype binaries between 0.7 and 1.9 M$_\odot$. Our target follows the distribution. According to the equations given by Y{\i}ld{\i}z (2014), we also estimated the age of our target as 5.3 Gyr from the initial mass of the two components (M$_{i1,2}$) and the mass derived from luminosity (M$_L$). The age includes detached and semidetached phases, and the phase from the beginning of contact phase to the present.

The orbital angular momentum (J$_o$) of our target was computed by using the following equation as Eker et al. (2006) did:
\begin{equation}
\ J_o = \frac{q}{(1+q)^2}\sqrt[3]{\frac{G^2}{2\pi}M^5P}, \\
\end{equation}
where M is the systemic mass and P is the orbital period. The relationship between logJ$_o$ and logM of NSVS 5029961 and the stars listed in Table 1 of the literature (Eker et al. 2006) is shown in Figure 9. The border line separating detached and contact systems was found by Eker et al. (2006). The value of logJ$_o$ = 51.57 is lower than the maximum of the orbital angular momentum of contact state for its total mass, suggesting that our target is in the contact phase.

Recently, Liu et al. (2018) tried to explain the relationship between q and f based on the contact binary samples from Yakut \& Eggleton (2005). Combining their explanation, we assume that the extremely low-mass-ratio contact binaries with deep and shallow degree of contact have different origins. The deep ones may evolve from high-mass-ratio contact binaries with the increasing of the fill-out factor, while the shallow ones may have low mass ratio at the beginning of contact phase after evolving from EA-type binaries through the AML via magnetic braking and case A mass transfer (Qian et al. 2018). Liu et al. (2018) also proposed a f-dominated simplification model for CCE-dominated mechanism. Before the two components merge, the mass ratio is gradually decreasing and the fill-out factor is oscillating. It is possible that the shallow and low-mass-ratio contact binaries were observed when they were at a state with small fill-out factor in one cycle of the value of the fill-out factor.

In order to discuss the stability of our target further, we calculated the instability parameters using a series of formulas introduced by Wadhwa et al. (2021). The instability mass ratio (q$_{inst}$ = 0.0342) was determined based on the mass of primary star and the fill-out factor. A new theoretical fractional radius of the primary was estimated to be 0.5534 from Equation (12) in the literature. Then the radius of the primary (R$_1$ = 1.570 R$_\odot$) was derived with the current separation. We set the value of k$_2$ as Wadhwa et al. (2021) did. The dimensionless gyration radius of the primary (k$_1$ = 0.1774) was estimated from Landin et al. (2009) for a 1.80 M$_\odot$ star. Using the values of R$_1$, k$_1$, k$_2$ mentioned above and photometric mass ratio, the instability separation (A$_{inst}$) can be determined to be 1.802 R$_\odot$. Combining q$_{inst}$ and A$_{inst}$, we expect that the period would be 0.1926 d when the instability occurs. The three instability parameters are smaller than the corresponding actual parameters, suggesting that the target is a relatively stable binary.

In conclusion, we analyzed the light curves and spectra of NSVS 5029961. The results show that it is an A-subtype shallow contact binary system with a low mass ratio and strong chromosphere activity. We corrected the period, determined absolute parameters from Gaia distance, and estimated the initial mass of the two components and the age of our target. The primary component is a little evolved star, while the secondary component has evolved away from the main sequence. It is supposed that the target may have evolved from a detached binaries with short period and low mass ratio by AML and be in a stable contact stage at present. More observations are necessary to help us understand the variation of orbital period, the formation and evolution of NSVS 5029961.

\begin{table}
\centering
\caption{Absolute Parameters of NSVS 5029961}
\begin{tabular}{p{2.7cm}p{2cm}p{2cm}}
\hline
Parameters           & Values          &Error \\
\hline
D(pc)                & 426.4           &5.9\\
m$_{Vmax}$(mag)      & 11.496          &0.015\\
m$_{V1}$(mag)        & 11.673          &0.017\\
m$_{V2}$(mag)        & 13.550          &0.016\\
M$_{V1}$(mag)        & 3.480           &0.047\\
M$_{V2}$(mag)        & 5.357           &0.046\\
BC$_{V1}$(mag)       & -0.060          &-  \\
BC$_{V2}$(mag)       & -0.070          &-  \\
M$_{bol1}$(mag)      & 3.420           &0.047 \\
M$_{bol2}$(mag)      & 5.287           &0.046 \\
L$_1$(L$_\odot$)     & 3.403           &0.147\\
L$_2$(L$_\odot$)     & 0.610           &0.026\\
R$_1$(R$_\odot$)     & 1.573           &0.119\\
R$_2$(R$_\odot$)     & 0.680           &0.053\\
a(R$_\odot$)         & 2.837           &0.236\\
M$_1$(M$_\odot$)     & 1.872           &0.468\\
M$_2$(M$_\odot$)     & 0.284           &0.072\\
\hline
\end{tabular}
\end{table}

\renewcommand\arraystretch{1.3}
\begin{table}
\tiny
\begin{center}
\caption{The shallow(f$\leq$25$\%$) and low-mass-ratio(q$\leq$0.25) contact binaries}
\resizebox{\textwidth}{!}
{
\begin{tabular}{p{2.9cm}p{0.9cm}p{0.5cm}p{0.5cm}p{0.6cm}p{0.7cm}p{0.6cm}p{0.6cm}p{0.7cm}p{0.7cm}p{0.6cm}p{0.6cm}p{0.6cm}p{0.6cm}p{1.8cm}p{1.3cm}p{1.7cm}}
\hline
Star                  &P      &T1  &T2  &q    &V    &f    &a        &M1   &M2   &L1      &L2      &R1   &R2   &dp/dt &J$_{spin}$/J$_{orb}$ & Ref. \\
 &(days)&(K) &(K) & &(mag) &(\%)  &(R$_\odot$)&(M$_\odot$) & (M$_\odot$) & (L$_\odot$)& (L$_\odot$)& (R$_\odot$)&(R$_\odot$)&($\times$10$^{-8}$d yr$^{-1}$) & & \\
\hline
ASAS J142124+1813.1   &0.24272&6160&5630&0.144&12.88&23   &2.40     &2.73 &0.39 &2.64    &0.29    &1.43 &0.57 &-     &0.153&(1), (2)\\
GV Leo                &0.26673&4850&5344&0.188&11.58&17.74&1.85$^a$ &1.01 &0.19 &0.50$^b$&0.18$^b$&1.01 &0.49 &-49.5 &0.113&(3), (4)\\
NSVS 7328383          &0.27208&5146&5056&0.205&13.01&17.0 &1.89$^a$ &1.01 &0.21 &0.63    &0.14    &1.00 &0.49 &-     &0.104&(5)\\
V345 Gem              &0.27477&6200&5840&0.171&8.03 &11   &2.07$^a$ &1.37 &0.20 &1.71    &0.25    &1.13 &0.49 &+5.88 &0.128&(6), (7)\\
PY Boo                &0.27805&4770&5080&0.222&12.40&17.8 &1.77     &0.79 &0.17 &0.40    &0.13    &0.93 &0.47 &-71.86&0.097&(1), (5)\\
ASAS J102556+2049.3   &0.28498&6010&5941&0.131&10.66&24.3 &1.69     &0.71 &0.09 &1.32    &0.17    &1.06 &0.39 &-339  &0.170&(1), (8)\\
UW CVn                &0.29247&5500&5889&0.245&14.85&20   &-        &-    &-    &-       &-       &-    &-    &-7.53 &0.086&(9), (10)\\
FU Dra                &0.30672&5800&6133&0.248&10.68&24   &2.17     &1.17 &0.29 &1.29$^b$&0.49$^b$&1.13 &0.62 &-     &0.086&(11)\\
NSVS 1917038          &0.31807&5074&4869&0.146&13.28&4    &-        &-    &-    &-       &-       &-    &-    &-     &0.147&(12)\\
GM Dra                &0.33875&6306&6450&0.180&8.80 &23   &2.33     &1.25 &0.22 &2.19    &0.56    &1.25 &0.61 &-     &0.118&(13)\\
NSVS 2256852          &0.34889&6804&5606&0.162&12.93&17   &2.16     &0.95 &0.15 &2.67    &0.24    &1.18 &0.52 &-23.6 &0.133&(1), (14)\\
NSVS 5029961          &0.37667&6260&6192&0.152&11.54&19.1 &2.84     &1.87 &0.28 &3.40    &0.61    &1.57 &0.68 &-     &0.143&(15)\\
USNO-B1.0 1452-0049820&0.37815&6400&6300&0.111&14.67&4.5  &2.50$^a$ &1.32 &0.15 &3.03    &0.39    &1.43 &0.53 &-     &0.199&(8)\\
GSC 03950-00707       &0.41198&5729&5623&0.205&13.36&11.83&3.52     &2.85 &0.58 &3.29    &0.74    &1.85 &0.91 &-     &0.102&(1), (16)\\
V2357 Oph             &0.41557&5640&5780&0.231&10.51&23   &2.67     &1.19 &0.29 &1.78    &0.47    &1.39 &0.69 &-     &0.091&(17), (18)\\
HI Pup                &0.43262&6500&6377&0.206&10.54&20   &2.70     &1.21 &0.23 &3.30    &0.70    &1.44 &0.67 &-     &0.103&(19)\\
V972 Her              &0.44309&6099&6522&0.167&6.65 &1    &2.49     &0.91 &0.15 &2.11    &0.57    &1.35 &0.59 &-     &0.127&(20), (21)\\
VV Cet                &0.52239&8100&5900&0.249&10.54&15.6 &3.75     &2.10 &0.52 &23.34   &1.00    &1.90 &0.99 &-     &0.082&(22)\\
eps CrA               &0.59144&6678&6341&0.128&5.61 &25   &3.69     &1.70 &0.23 &7.75    &1.02    &2.10 &0.85 &+46.7 &0.175&(23), (24)\\
HI Dra                &0.59742&7220&6890&0.250&9.01 &24   &3.86     &1.72 &0.43 &9.60    &2.40    &1.98 &1.08 &-     &0.085&(25), (26), (27)\\
V402 Aur              &0.60350&6775&6700&0.201&8.95 &3    &3.77     &1.64 &0.33 &7.43    &1.49    &2.00 &0.92 &-     &0.103&(28), (29)\\
V144 NGC 6715         &0.72159&7120&6808&0.160&19.27&19.5 &3.89     &1.31 &0.21 &10.19   &1.69    &2.23 &0.95 &-     &0.135&(30)\\
V2787 Ori             &0.81098&6993&5418&0.120&12.7 &18.1 &4.29     &1.44 &0.17 &12.93   &0.71    &2.45 &0.96 &-     &0.186&(31)\\
V921 Her              &0.87738&7780&7346&0.226&9.49 &23   &5.29     &2.07 &0.51 &23.50   &5.09    &2.75 &1.41 &-     &0.092&(17), (18)\\
NS Cam                &0.90739&6250&5689&0.213&12.78&17   &4.45     &1.18 &0.25 &7.64$^b$&1.27$^b$&2.36 &1.16 &-22.5 &0.099&(32)\\
\hline
\end{tabular}
}
\end{center}
(1) Kjurkchieva et al. (2019a); (2) Michel \& Kjurkchieva (2019); (3) Kriwattanawong \& Poojon (2013); (4) Zhang \& Qian (2020); (5) Kjurkchieva et al. (2019b); (6) Gazeas et al. (2021); (7) Yang et al. (2009); (8) Kjurkchieva et al. (2018a); (9) Kopacki \& Pigulski (1995); (10) Qian (2001); (11) Va$\check{n}$ko et al. (2001); (12) Guo et al. (2020); (13) Gazeas et al. (2005); (14) Kjurkchieva et al. (2018b); (15) This paper; (16) Kjurkchieva et al. (2018c); (17) Rucinski et al. (2003); (18) Gazeas et al. (2006); (19) Ula\c{s} \& Ulusoy (2014); (20) Selam et al. (2018); (21) Rucinski et al. (2002); (22) Rahman (2000); (23) Yang et al. (2005); (24) Goecking \& Duerbeck (1993); (25) Pribulla et al. (2009); (26) \c{C}al{\i}\c{s}kan et al. (2014); (27) Papageorgiou \& Christopoulou (2015); (28) Zola et al. (2004); (29) Pych et al. (2004); (30) Li \& Qian (2013); (31) Tian et al. (2019); (32) Samec et al. (2020).

$^a$ The value of a was calculated by us with Kepler's third law.

$^b$ The value of L was calculated by us with the the Stefan-Boltzmann law.
\end{table}

\begin{figure}\centering
\includegraphics[width=0.52\textwidth]{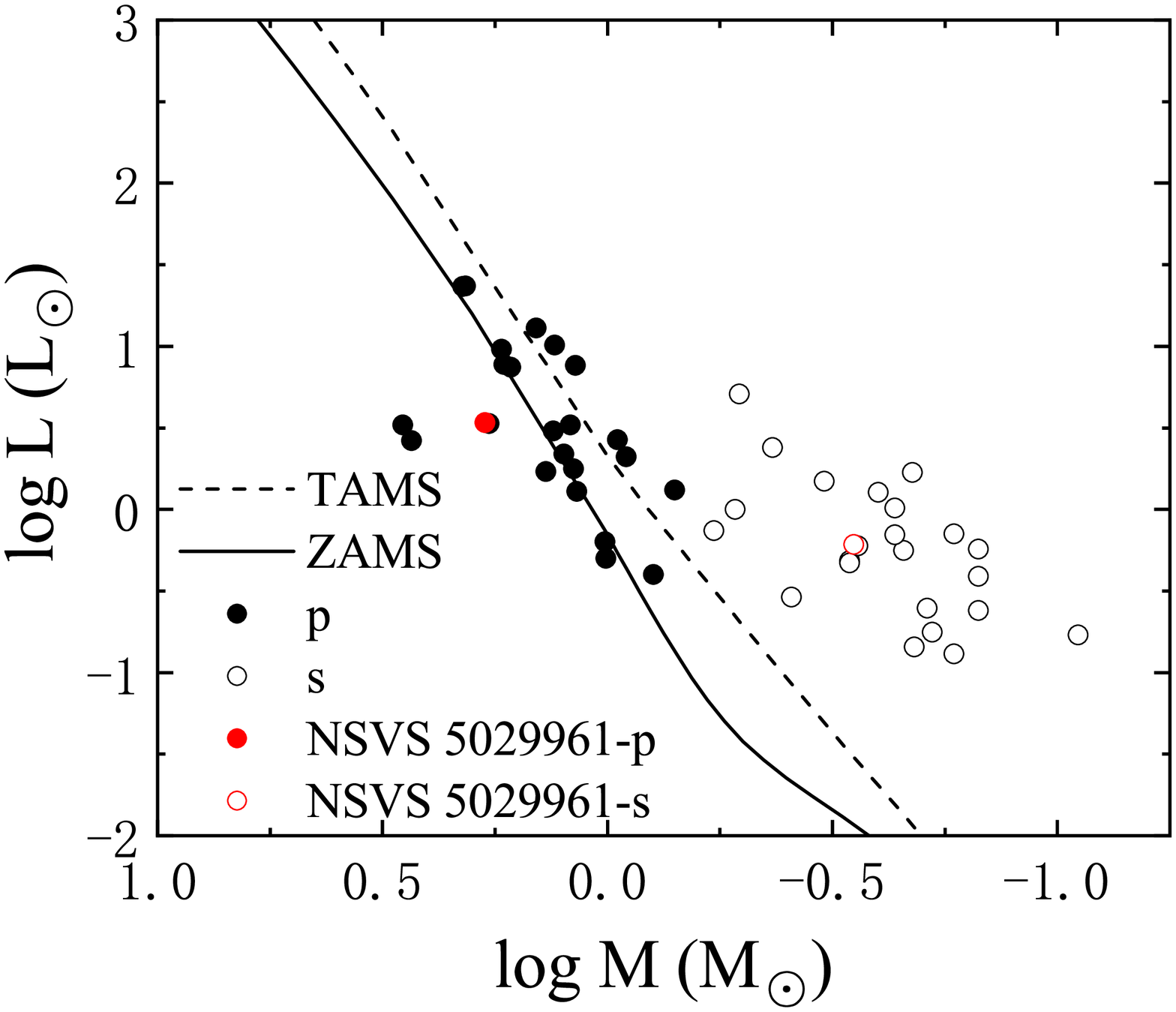}
\hspace{-1.5cm}
\includegraphics[width=0.52\textwidth]{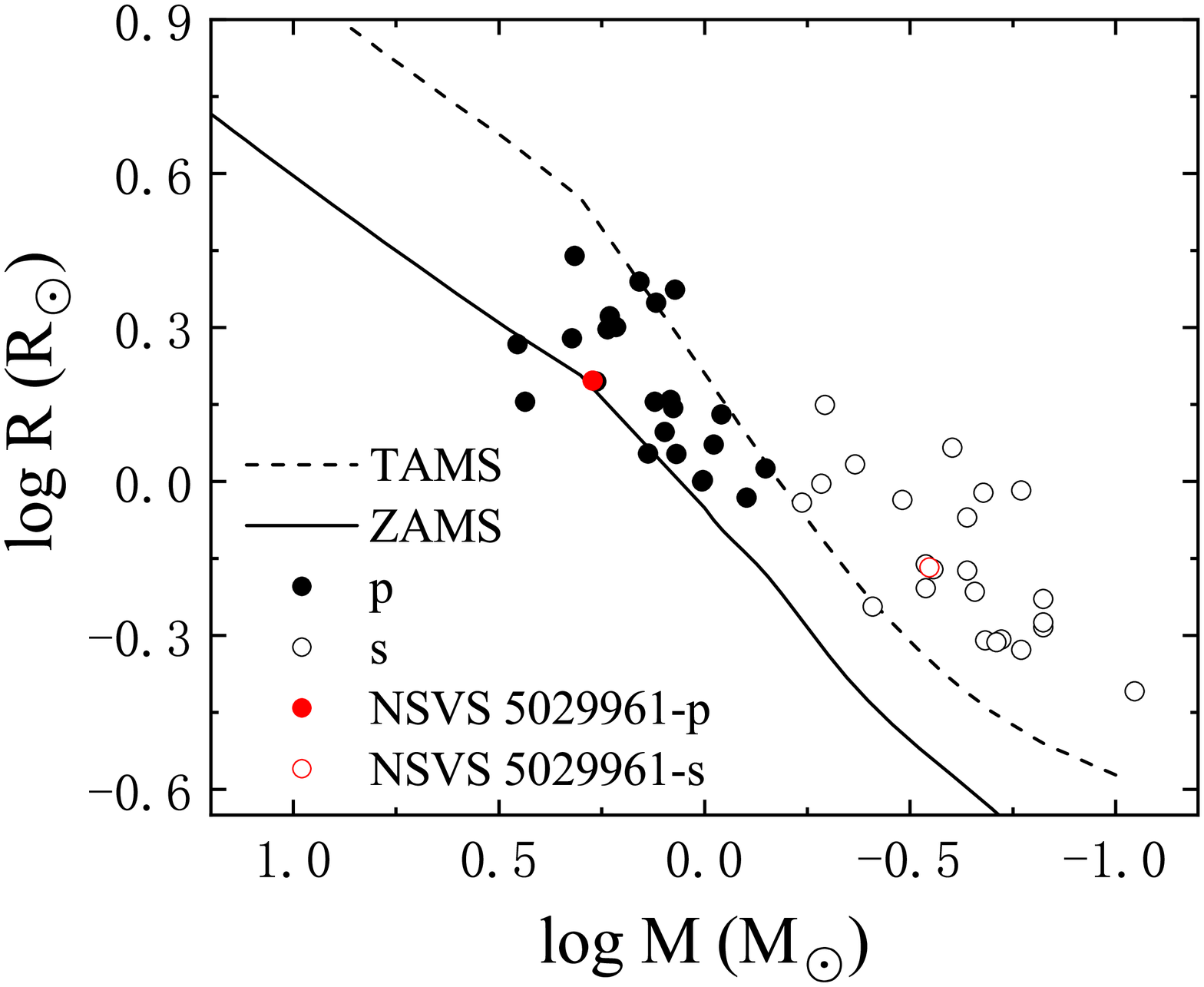}
\caption{The mass-luminosity diagram (left) and the mass-radius diagram (right).}
\end{figure}

\begin{figure}\centering
\includegraphics[width=0.6\textwidth]{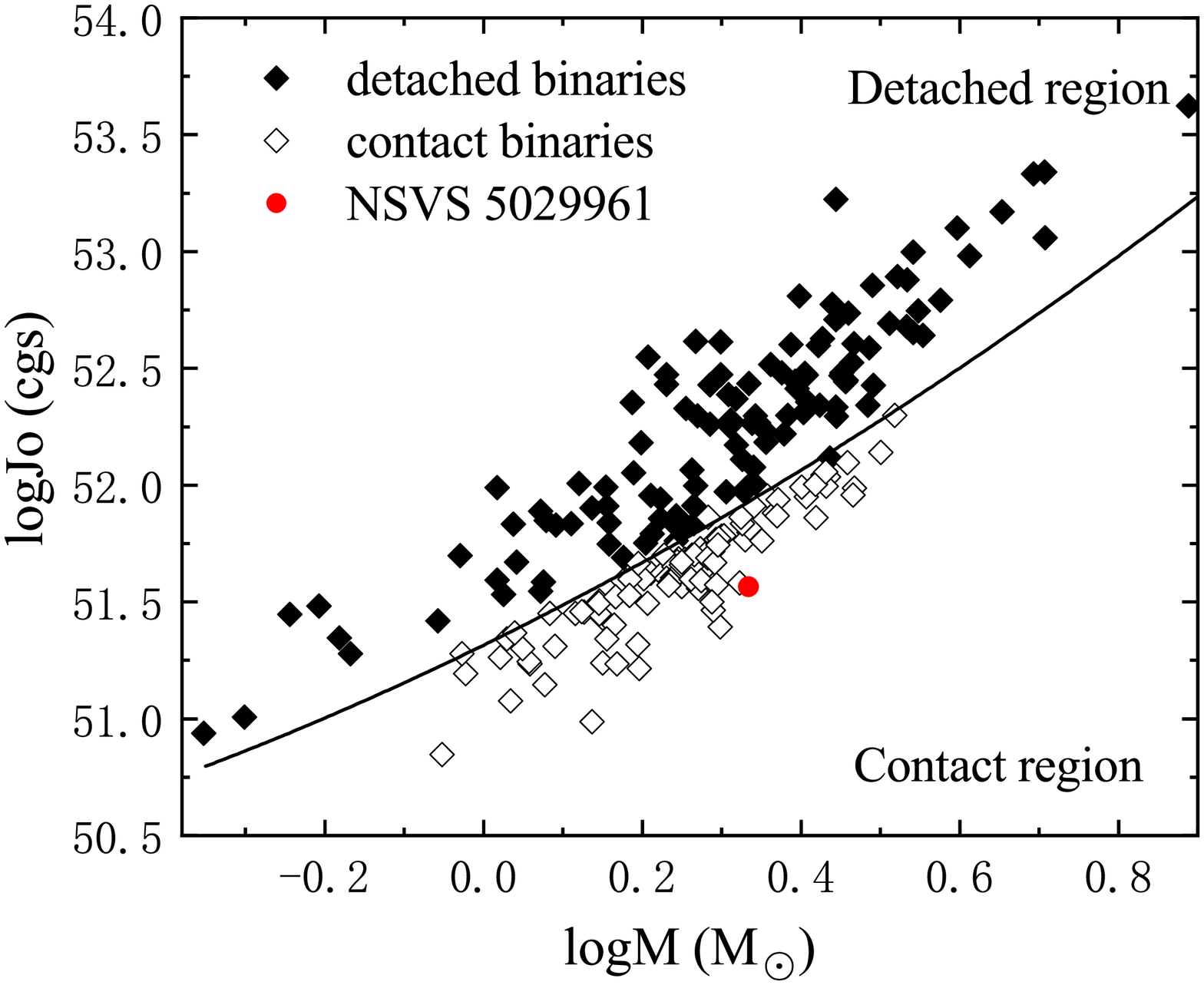}
\caption{The logJ$_o$-logM diagram, where the solid diamonds refer to the detached binaries, and the open ones refer to the contact binaries in the Table 1 of the literature (Eker et al. 2006). The red circle represents our target.}
\end{figure}

\section*{Acknowledgments}
Great thanks to the referee for very helpful comments and suggestions that improved our manuscript a lot. This work is supported by the Joint Research Fund in Astronomy (No. U1931103) under cooperative agreement between National Natural Science Foundation of China (NSFC) and Chinese Academy of Sciences (CAS), and by NSFC (No. 11703016), and by the Natural Science Foundation of Shandong Province (No. ZR2014AQ019), and by Young Scholars Program of Shandong University, Weihai (No. 20820171006), and by the Open Research Program of Key Laboratory for the Structure and Evolution of Celestial Objects (No. OP201704).

This work is partly supported by the Supercomputing Center of Shandong University, Weihai.

The spectral data were provided by Guoshoujing Telescope (the Large Sky Area Multi-Object Fiber Spectroscopic Telescope LAMOST), which is a National Major Scientific Project built by the Chinese Academy of Sciences. Funding for the project has been provided by the National Development and Reform Commission. LAMOST is operated and managed by the National Astronomical Observatories, Chinese Academy of Sciences.

This work has made use of data from the European Space Agency (ESA) mission
{\it Gaia} (\url{https://www.cosmos.esa.int/gaia}), processed by the {\it Gaia}
Data Processing and Analysis Consortium (DPAC,
\url{https://www.cosmos.esa.int/web/gaia/dpac/consortium}). Funding for the DPAC
has been provided by national institutions, in particular the institutions
participating in the {\it Gaia} Multilateral Agreement.

This work includes data collected by the TESS
mission. Funding for the TESS mission is provided by NASA Science Mission directorate. We
acknowledge the TESS team for its support of this work.

This publication makes use of data products from the AAVSO
Photometric All Sky Survey (APASS). Funded by the Robert Martin Ayers
Sciences Fund and the National Science Foundation.

This publication makes use of data products from the Two Micron All Sky Survey, which is a joint project of the University of Massachusetts and the Infrared Processing and Analysis Center/California Institute of Technology, funded by the National Aeronautics and Space Administration and the National Science Foundation.

This paper makes use of data from the DR1 of the WASP data (\citealt{Butters2010}) as provided by the WASP consortium,
and the computing and storage facilities at the CERIT Scientific Cloud, reg. no. CZ.1.05/3.2.00/08.0144
which is operated by Masaryk University, Czech Republic.

\section*{Data availability}
The TESS data are publicly available at http://archive.stsci.edu/tess/bulk\_downloads.html, SuperWASP data are publicly available at https://wasp.cerit-sc.cz/search, ASAS-SN data are publicly available at https://asas-sn.osu.edu/variables/lookup, and NSVS data are publicly available at https://skydot.lanl.gov/nsvs/cone\_search.php?ra=189.03725\&dec=42.24303\&rad=0.5\&saturated=on\&apincompl=on\&nocorr=on\&hiscat=on\&hicorr=on.
The WHOT data are available in the machine-readable form of Table 2.

\end{document}